%% file: spdbl-mloss.tex
\documentclass[twoside,11pt]{article}

\usepackage{float}
\usepackage[preprint]{jmlr2e}
\usepackage[T1]{fontenc}
\usepackage{lmodern}
\usepackage{booktabs}
\usepackage{amsmath}
\usepackage{bm}
\usepackage{xcolor}
\usepackage{fancyvrb}
\usepackage{algorithm}
\usepackage{algpseudocode}

\newcommand{\pkg}[1]{\textsf{#1}}
\newcommand{\code}[1]{\texttt{#1}}
\algrenewcommand\algorithmicrequire{\textbf{Input:}}
\algrenewcommand\algorithmicensure{\textbf{Output:}}

\floatstyle{plaintop}
\newfloat{listing}{t}{lol}
\floatname{listing}{Listing}

\hypersetup{pdftitle={Dynamic Emulation for Bayesian Inverse Problems in
  Spatiotemporal Mechanistic Models}, pdfauthor={Xiang Chen and Sudipto Banerjee}}

\jmlrheading{}{2026}{}{}{}{}{Xiang Chen and Sudipto Banerjee}
\ShortHeadings{Dynamic Emulation for Bayesian Inverse Problems}{Chen and Banerjee}
\firstpageno{1}

\begin{document}

\title{Dynamic Emulation for Bayesian Inverse Problems\\ in Spatiotemporal Mechanistic Models}

\author{\name Xiang Chen \email xiangchen@ucla.edu \\
       \name Sudipto Banerjee \email sudipto@ucla.edu \\
       \addr Department of Biostatistics,
       University of California, Los Angeles,
       Los Angeles, CA 90095, USA}

\editor{}

\maketitle

\begin{abstract}%
Calibrating a spatiotemporal mechanistic model to field data is a Bayesian
inverse problem in which every likelihood evaluation needs a simulator run.
The R package \pkg{spDBL} replaces the simulator with a dynamic emulator, a
matrix-variate dynamic linear model with a closed-form posterior that evolves
the field in discrete time and processes large grids in spatial blocks. It then
calibrates the parameters without further simulator runs, using Stan
conditionally on emulator draws. In benchmarks on epidemic fields, with every
method tuned alike, \pkg{spDBL}'s predictions at untried inputs were the only
ones with near-nominal interval coverage, though a tuned \pkg{RobustGaSP} was
more accurate, and \pkg{spDBL}'s calibrations matched the accuracy of
\pkg{RobustCalibration} with higher, though below nominal, interval coverage
(80\% against 67--76\%) and over 50 times more effective samples per CPU
second. The MIT-licensed package is on CRAN and at
\url{https://github.com/xiangchen-stat/spDBL}.
\end{abstract}


\section{Introduction}

Mechanistic or physics-driven models such as reaction--diffusion systems advance a spatial field in discrete time steps for every setting of their inputs. \pkg{spDBL} addresses two problems they pose:
\emph{discrete-time modelling}, in which a statistical \emph{emulator} trained on a modest design of runs models each time step's field given the previous ones, as the simulator does; and the \emph{Bayesian inverse problem} of inferring the inputs from noisy field observations \citep{Stuart2010}, in which the emulator stands in for the costly simulator and \emph{calibration} accounts for its error and the model's discrepancy from reality \citep{Kennedy2001}. Existing software treats the space--time output as one long vector with a shared Gaussian process \citep{Gu2016,Gu2019}, as a separable process over inputs and time \citep{stilt}, or through a principal-component basis \citep{Higdon2008,SEPIA,surmise}; other emulation packages \citep{Hankin2005,Palomo2015,Gramacy2016,Francom2020,Gu2023}
have no dynamic model, and dynamic emulators \citep{LiuWest2009,Conti2010,dynemu} and dynamic linear model and Kalman-filter Gaussian process software \citep{Petris2010,Helske2017,Finley2015,FastGaSP} handle time series or observed data.
While fast and accurate for their purposes (especially multiple packages from Mengyang Gu and collaborators), \pkg{spDBL} is distinct in its scalable implementation of \citet{Banerjee2025} as reusable software and, beyond its research code (see \code{IMPROVEMENTS.md}),
calibrates without simulator runs, with the No-U-Turn sampler (NUTS;
\citealp{Hoffman2014}), cached compiled models, resumable and parallel fits, and tests against independent calculations.

\section{Models and Computation}

Each run's output at time $t$ is a field over $S$ sites, split into blocks of $C$ sites ($w$ grid columns, the \emph{episode window}); each (time, block) pair is an \emph{episode} $e$. With $\bm Y_e$ the $N\times C$ outputs of $N$ training runs and $\bm F_e$ the $N\times p$ design built from their two previous
outputs ($p=2C$), the emulator follows \citet[Equation 2]{Banerjee2025},
\begin{equation}\label{eq:emulator}
\begin{split}
\bm Y_e &= \bm F_e\bm\Theta_e + \bm{\mathcal{E}}_e,\quad
  \bm{\mathcal{E}}_e\stackrel{\text{ind.}}{\sim}\mathcal{MN}_{N\times C}(\bm O, \bm V, \bm\Sigma),\\
\bm\Theta_e &= \bm G\bm\Theta_{e-1} + \bm\Gamma_e,\quad
  \bm\Gamma_e\stackrel{\text{ind.}}{\sim}\mathcal{MN}_{p\times C}(\bm O, \bm W, \bm\Sigma),
  \quad e = 1, 2, \ldots, E,\\
\bm\Theta_0\mid\bm\Sigma &\sim \mathcal{MN}_{p\times C}(\bm m_0, \bm M_0, \bm\Sigma),\quad
  \bm\Sigma\sim\mathcal{IW}(n_0, \bm D_0)\ \text{ or }\ \bm\Sigma = \sigma^2\bm R,\
  \sigma^2\sim\mathcal{IG}(n_0, d_0),
\end{split}
\end{equation}
where $\bm V$ is a Gaussian-process covariance over the inputs, $\bm G = \bm W =
\bm I$ and $\bm R$ is a spatial correlation. Forward filtering and backward
smoothing \citep{Carter1994,Fruhwirth1994,West1997} give each episode's
posterior in closed form. Field data at observed sites $\bm s$ follow
\citet[Equation 27]{Banerjee2025},
\begin{equation}\label{eq:calibration}
\begin{split}
z_t(\bm s) &= y_t(\bm\eta, \bm s) + u_t(\bm s) + \varepsilon^z_t(\bm s),\quad
  \varepsilon^z_t(\bm s)\stackrel{\text{ind.}}{\sim}\mathcal{N}(0, \tau_t^2),\\
u_t(\bm s) &= u_{t-1}(\bm s) + \varepsilon^u_t(\bm s),\quad
  \varepsilon^u_t(\bm s)\stackrel{\text{ind.}}{\sim}\mathcal{GP}(0, \tau_t^2\,\mathcal{C}(\cdot,\cdot)),\\
\tau_t^2 &\stackrel{\text{ind.}}{\sim}\mathcal{IG}(b^t n^z_0, b^t d^z_0),\quad t = 1, 2, \ldots, T,\quad
  \bm\eta\sim p(\bm\eta),
\end{split}
\end{equation}
where $p(\bm\eta)$ is uniform on a box and $\bm U$, the matrix of $\mathcal{C}$
at the observed sites, is user-supplied. At an untried $\bm\eta$ a Gaussian process (Mat\'ern 5/2, one per
time step) predicts $\bm F(\bm\eta)$ with its uncertainty. Given one emulator
draw, $y_t$ and $u_t$ are integrated out (the latter by a Kalman filter) and
Stan \citep{Carpenter2017} samples $(\bm\eta,\bm\tau^2)$ with NUTS from the best
of several local optima; pooling over $M$ draws gives a modular (``cut'')
posterior \citep{Liu2009,Plummer2015}. Appendix~\ref{app:prediction} derives
predictions at unobserved sites.

\section{Software Design and Usage}

\begin{figure}[t]
\centering
\includegraphics[width=0.95\linewidth]{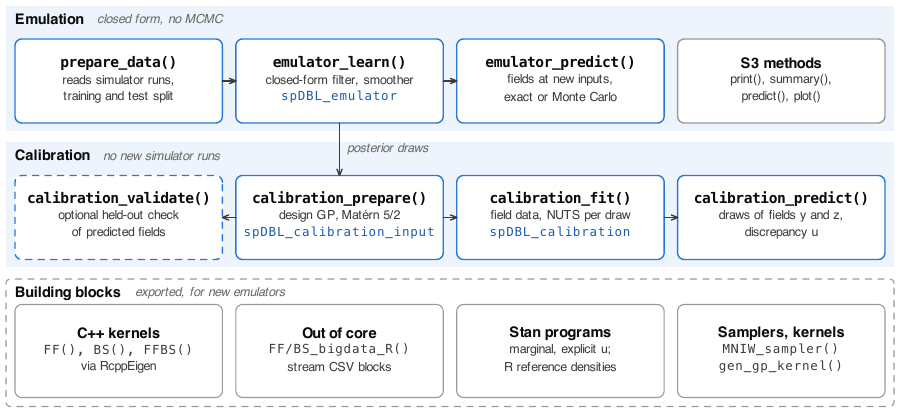}
\caption{Architecture of \pkg{spDBL}. Blue: workflow functions and classes; dashed: optional.}
\label{fig:architecture}
\end{figure}

Seven workflow functions cover emulation and calibration
(Figure~\ref{fig:architecture}); the underlying filters, smoothers, samplers
and predictors are exported as building blocks alongside closed-form WAIC
scores (Appendix~\ref{app:waic}), with C++ inverse-Wishart steps
(\pkg{RcppEigen}; \citealp{Eddelbuettel2013}), out-of-core routines for
oversized grids and stored or regenerable coefficient draws. Results are
serializable S3 objects recording settings, seed and version.
Listing~\ref{lst:calibration} calibrates a spatial
susceptible--infected--recovered (SIR) model~\eqref{eq:sir} from 40 shipped
runs on a $6\times6$ grid, initialized at the best of 20 L-BFGS optima. On 20
cores it runs in 26 minutes with $\hat R\le 1.007$ \citep{Vehtari2021}, bulk
and tail ESS $\ge 1{,}835$ and $738$, no divergences and 95\% intervals
covering the truth.

\begin{listing}[t]
\caption{Calibration from saved simulations (\code{benchmarks/listing1.R} in the repository).}
\label{lst:calibration}
\begin{Verbatim}[fontsize=\scriptsize, frame=lines, framesep=1mm]
library(spDBL)
data("dt_calibration"); d <- dt_calibration
em  <- emulator_learn(d$training$input, d$training$responses, d$Nx, d$Ny,
                      covariance = "sigma2R", episode_window = 2,
                      nsam = 20, seed = 1)               # closed form, no MCMC
cal <- calibration_prepare(em, d$lower, d$upper)        # design GP
calibration_validate(cal, d$validation$input, d$validation$responses)$field
fit <- calibration_fit(cal, d$field$z, d$field$U, chains = 4, iter = 2000,
  parallel_draws = 20, seed = 1)                        # NUTS, 20 cores
summary(fit)                                            # eta, R-hat, ESS
fields <- calibration_predict(fit)                      # y, u and z draws
\end{Verbatim}
\end{listing}

\section{Evaluation}

\begin{table}[t]
\centering\scriptsize\setlength{\tabcolsep}{4pt}
\begin{tabular}{@{}lccccc@{}}
\toprule
 & Spatial & Dynamic & Emulator & Calibration with & Large \\
Software & fields & emulator & without MCMC & discrepancy & outputs \\
\midrule
\pkg{spDBL} & yes & yes & yes & dynamic (NUTS) & blocks \\
Research code$^{a}$ & yes & yes & yes & dynamic (MCMC) & blocks \\
\pkg{RobustGaSP} (\code{ppgasp}) & as vector & no & yes & via \pkg{RobustCalibration} & shared kernel \\
\pkg{RobustCalibration} & as vector & no & yes & yes (MCMC) & shared kernel \\
\pkg{SEPIA} / \pkg{GPMSA} & PCA basis & no & no & yes (MCMC) & PCA \\
\pkg{surmise} & PCA basis & no & yes & no$^{b}$ & PCA \\
\pkg{dynemu} & no & yes & yes & no & -- \\
\bottomrule
\end{tabular}
\caption{Features of \pkg{spDBL} and related software (November 2026).
$^{a}$\citet{Banerjee2025}, solving the simulator at every MCMC step.
$^{b}$MCMC without a discrepancy.}
\label{tab:features}
\end{table}

\begin{figure}[t]
\centering
\includegraphics[width=0.95\linewidth]{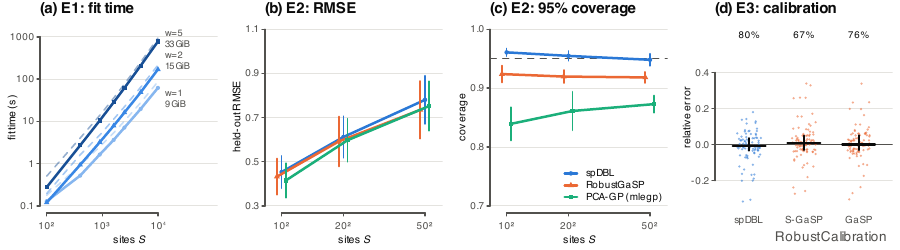}
\caption{(a)~E1: fit time by window $w$ (dashed: stored draws), peak
memory at $S=10^4$. (b,~c)~E2: RMSE and 95\% coverage (mean, 95\% interval, 10
designs). (d)~E3: error relative to prior range in 30 calibrations (median,
quartiles), 95\% coverage.}
\label{fig:benchmarks}
\end{figure}

We compared \pkg{spDBL} (Table~\ref{tab:features}) with \pkg{RobustGaSP}
\citep{Gu2019}, a principal-component Gaussian process (PCA-GP) in
\pkg{mlegp} \citep{mlegp}, and \pkg{RobustCalibration} \citep{Gu2023} with
S-GaSP and GaSP discrepancies \citep{GuWang2018}, on SIR fields~\eqref{eq:sir}
from 40 training runs (E3: 100; Figure~\ref{fig:benchmarks};
Appendix~\ref{app:benchmarks}). All methods were tuned alike.

\paragraph{Scaling (E1).}
With $10^4$ sites and regenerated draws, fitting took 1.0~min and 9~GiB (window
1) to 12.9~min and 33~GiB (window 5; 62~GiB if stored), growing as $S^{1.4}$--$S^{1.7}$.

\paragraph{Emulation at untried inputs (E2).}
On one $10\times10$ domain at $10^2$, $20^2$ and $50^2$ sites, with 10 designs
of 40 training and 50 held-out runs, only \pkg{spDBL}'s
predictions had near-nominal 95\% intervals (94.8--96.1\%
coverage against 91.8--92.4\% for \pkg{RobustGaSP} and 83.9--87.3\% for the
PCA-GP). \pkg{RobustGaSP} was more accurate: its CRPS \citep{Gneiting2007} was lower on every grid and
its RMSE lower, significantly at $50^2$ (0.735 against 0.780); the PCA-GP's
did not differ significantly from \pkg{spDBL}'s.

\paragraph{Calibration (E3).}
In 30 calibrations of three parameters from 100 training runs,
\pkg{spDBL} matched \pkg{RobustCalibration}'s accuracy (relative RMSE 0.087
against 0.103 and 0.092 for S-GaSP and GaSP), and its 95\% intervals covered
80\%, against 67\% for S-GaSP (a significant difference) and 76\% for GaSP.
Chains mixed ($\hat R \le 1.1$) in all but one replicate per method, and
\pkg{spDBL} gave 72 and 54 times more effective samples per CPU second. With 40 bundled runs every method
under-covered, \pkg{spDBL}'s most (54\%; Appendix~\ref{app:benchmarks}).

\section{Availability and Limitations}\label{sec:limitations}

\pkg{spDBL} 1.2.0 (tag \code{v1.2.0},
\href{https://doi.org/10.32614/CRAN.package.spDBL}{doi:10.32614/CRAN.package.spDBL})
is MIT-licensed on CRAN, with Linux, macOS and Windows CI and 99.1\% R/C++
test coverage checking each numerical component independently; the website
hosts vignettes and an architecture guide, and the repository provides a
contributing guide, issue tracker and forum. Outputs require complete
single-grid fields, prediction is costlier than competitors'
(Table~\ref{tab:e2}), and calibration requires a user-supplied $\bm U$, costs
$O(T S_o^3)$ for $S_o$ observed sites and under-covers intervals from emulator
error (Appendix~\ref{app:benchmarks}). Rather than fixing GP hyperparameters,
future versions will stack closed-form posteriors over a grid by
cross-validation \citep[predictive stacking;][]{Zhang2026,Pan2025}.

\acks{
The authors acknowledge funding from the National Science Foundation through DMS-1916349 and DMS-2113778, and from the National Institutes of Health through NIEHS-R01ES027027, NIEHS-R01ES030210 and NIGMS-R01GM148761. The authors declare no competing interests.}

\bibliography{refs}

\clearpage
\makeatletter\setlength{\@fptop}{0pt}\makeatother
\appendix
\renewcommand{\thetable}{A\arabic{table}}
\renewcommand{\theHtable}{A\arabic{table}}
\setcounter{table}{0}
\section{Benchmark Results}\label{app:benchmarks}

The tables at the end of this appendix give every metric behind
Figure~\ref{fig:benchmarks}. They are generated from the committed results by
\code{benchmarks/paper/tables.R}; \code{benchmarks/README.md} maps each result
folder to this paper.

\paragraph{Simulator.} Listing~\ref{lst:calibration} and E1--E3 use the
spatial SIR model of \citet[Equation~26]{Banerjee2025}, in which the numbers
of susceptible, infected and recovered people, $S_t(\bm s)$, $I_t(\bm s)$ and
$R_t(\bm s)$, at site $\bm s$ evolve as
\begin{equation}\label{eq:sir}
\begin{split}
\frac{\partial S_t(\bm s)}{\partial t} &= -\eta_1\frac{S_t(\bm s)I_t(\bm s)}{N}
  + \alpha_1\nabla^2 S_t(\bm s),\\
\frac{\partial I_t(\bm s)}{\partial t} &= \eta_1\frac{S_t(\bm s)I_t(\bm s)}{N}
  - \eta_2 I_t(\bm s) + \alpha_2\nabla^2 I_t(\bm s),\\
\frac{\partial R_t(\bm s)}{\partial t} &= \eta_2 I_t(\bm s)
  + \alpha_3\nabla^2 R_t(\bm s),
\end{split}
\end{equation}
where $\eta_1$ is the transmission rate, $\eta_2$ the recovery rate,
$\alpha_1$, $\alpha_2$ and $\alpha_3$ are diffusion coefficients, $N$ is the
population of a grid cell (here, not the number of training runs) and
$\nabla^2 f = \partial^2 f/\partial s_1^2 + \partial^2 f/\partial s_2^2$.
\code{gen\_pde()} solves~\eqref{eq:sir} by the method of lines on unit grid
cells (\pkg{deSolve} and \pkg{ReacTran}; \citealp{deSolve,ReacTran}), with
values fixed at zero outside the grid (an absorbing boundary), $N = 10{,}000$
susceptible people in every cell, one infected person in the central cell and
$0.05N$ in a cell near a corner; E2 solves the same system on a fixed
$10\times10$ domain at every resolution. The emulated output is
$y_t(\bm\eta, \bm s) = \log\{I_t(\bm s) + 1\}$ at the solver's times $0, 1, \ldots, 25$,
indexed $t = 1, \ldots, 26$, with
$\eta_1\in[2, 4]$, $\eta_2\in[0.2, 0.4]$, $\alpha_2\in[0, 0.2]$ and
$\alpha_1 = \alpha_3 = 0$.

\paragraph{Settings.} One core per task on Intel Xeon Gold 6240 nodes (E3 with
40 training runs: 18 tasks per 36-core node; with 100: one Slurm job per
calibration, \pkg{RobustCalibration} running its four chains on four cores), R 4.6.1 with oneMKL; \pkg{spDBL} 1.2.0,
\pkg{RobustGaSP} 0.6.8, \pkg{RobustCalibration} 0.5.6, \pkg{mlegp} 3.1.10.
\pkg{RobustCalibration} ran 4 chains of 20{,}000 iterations (5{,}000 burn-in),
twice its default length, and \pkg{spDBL} 4 chains of 2{,}000 iterations for
each of 20 emulator draws. E3 with 40 runs
ran before two later speed-ups in 1.2.0: calibration gradients are now 1.3
times faster (the same log density, compiled with stanc's O1 optimisations),
and conditional fits can run in parallel (\code{parallel\_draws}). E3 with 100
runs has the faster gradients and runs the conditional fits in sequence.

\paragraph{Tuning.} Every method was tuned by the same procedure, on data the
evaluation does not use or by rules that use no true values. In E2 each
method's settings were compared on two further designs (11 and 12) at all
three resolutions: round 1 changed one setting at a time, round 2 combined the
changes that lowered CRPS, and the setting with the lowest mean CRPS was
selected, a rule fixed before any result (Table~\ref{tab:e2-tuning}). The selected settings are
\pkg{spDBL} with episode window 5, \code{gp\_tune = 1} and 50 draws (CRPS 1.5\%
lower than with its E2 setting), \code{ppgasp} fitted per time step with a
linear trend and the Mat\'ern 3/2 kernel (23\% lower), and the PCA-GP with a
linear mean, 99.99\% of variance, a nugget and 20 restarts (7.5\% lower). In
E3 both packages start their chains at their optimum from 20 starts
(\pkg{spDBL}: L-BFGS on each conditional posterior; \pkg{RobustCalibration}:
maximum likelihood), and \pkg{RobustCalibration}'s emulator estimates a nugget,
as its tuned E2 fits do. \pkg{spDBL}'s NUTS adapts its step size and mass
matrix during warm-up; \pkg{RobustCalibration}'s random-walk proposal for
$\bm\eta$ has a fixed scale, by default 5\% of each prior range, which its
sampler multiplies by the range again, so that its steps were 10\% of the
range for $\eta_1$ and 1\% for $\eta_2$ and $\alpha_2$ and accepted 11\% of
moves. Each scale is therefore set to 1.2 times the parameter's posterior
standard deviation in a 4{,}000-iteration pilot chain from the optimum; the
factor 1.2 gave the most effective samples per CPU second on replicates 1--3
among 0.7, 1.2 and 2.0. CPU times include the pilot. Results before tuning are
kept in the repository.

\paragraph{E1.} Each episode costs time cubic in the block size $C = wN_y$ for
a grid $N_y$ sites high, so on square grids the cost grows at most as
$T w^2 S^2$; the log--log slopes of fit time against $S$ were 1.3, 1.5 and 1.6
for windows 1, 2 and 5 with stored draws, and 1.4, 1.6 and 1.7 with regenerated
draws. With coefficient draws regenerated on demand
(\code{emulator\_learn(keep\_draws = FALSE)}), peak memory at $10^4$ sites fell
from 13, 23 and 62~GiB to 9, 15 and 33~GiB, fitting was 17--29\% faster, and
Monte Carlo prediction 3--8 times slower, with identical results
(Table~\ref{tab:e1}).

\paragraph{E2.} The grids discretise the same physical domain, so they
simulate the same epidemic at increasing resolution. \pkg{spDBL}'s intervals,
from \code{predict()} with \code{interval = "prediction"}, are moment-matched
Gaussians over its emulator draws. With every method tuned
(Tables~\ref{tab:e2} and~\ref{tab:e2-paired}), the competitors' coverage was
3--4 points (\pkg{RobustGaSP}) and 8--12 points (PCA-GP) lower than
\pkg{spDBL}'s, significantly on every grid; \pkg{RobustGaSP}'s CRPS was lower
by 0.019--0.047 on every grid and its RMSE by 0.020--0.045, significantly only at
$50^2$; the PCA-GP's RMSE and CRPS did not differ significantly from
\pkg{spDBL}'s. Tuning changed \pkg{spDBL} little and \pkg{RobustGaSP} most,
by fitting one Gaussian process per time step instead of one with range
parameters shared by all outputs. Before tuning, the
competitors' coverage was on average 6--14 points lower than \pkg{spDBL}'s, and
lower in 119 of the 120 comparisons on the same design; the PCA-GP's RMSE was
lower by 0.014--0.026, within design-to-design variation, and
\pkg{RobustGaSP}'s higher on every grid, by 0.10--0.14 at $50^2$ for its
default and per-time-step fits. The
design Gaussian process alone covers 94.6--94.9\%, with 1--1.5\% higher RMSE,
and has lower CRPS on the $10^2$ grid (by 0.003). When the autoregressive rows come from the simulator's
earlier outputs at the test input, one-step predictions have RMSE 0.07--0.22,
15--29\% of the simulator-free error. Times and memory in Table~\ref{tab:e2}
come from a rerun of design 1 with \pkg{spDBL} 1.2.0's batched prediction, which
predicts the 50 runs about 50 times faster than the version first benchmarked
(43, 133 and 905~s on the three grids) with identical results; with
regenerated draws, peak memory at $50^2$ is 2.4 instead of 5.3~GiB.

E2 replaced an earlier profile, kept in \code{benchmarks/results/e2/paper},
that had one design per grid and unit grid cells, so that, as in E1, the larger
grids were larger domains with different epidemics, which confounds resolution
with the problem; the fixed-domain profile was set before any of its results
were seen. In the earlier profile \pkg{spDBL}'s intervals also covered
95.0--95.6\% on every grid and the competitors' 76--91\% on the $10^2$ and
$20^2$ grids, but on the $50^2$ grid tuned \pkg{RobustGaSP} had lower RMSE
than \pkg{spDBL} (0.66 against 0.70) and similar coverage (95.4\% against
95.0\%).

\paragraph{E3.} With random starts (kept in \code{benchmarks/results/e3/paper}),
$\hat R$ exceeded 1.1 in 26 of 30 replicates, and 24 \pkg{spDBL} chains on one
replicate and emulator draw found six modes: one near the true parameters and
five 36 to 586 log-density units below it, three of them on the boundary of the
prior box. Because a conditional fit's chains now
start near one optimum, $\hat R$ checks mixing within its mode; the fit records
the optima of all 20 starts. $\hat R$ is rank-normalised, and reported for the
worst of \pkg{spDBL}'s 20 conditional fits. Effective sample sizes per CPU
second are summed over the conditional fits, each keeping all 4{,}000
post-warm-up draws (the default keeps 200), and \pkg{RobustCalibration}'s
over its four chains.

With 100 training runs (one Latin-hypercube design, the same for every
replicate and method; Table~\ref{tab:e3}), a paired replicate bootstrap finds
no accuracy difference between the methods. \pkg{spDBL}'s intervals covered
0.13 [0.01, 0.26] more than tuned S-GaSP's, mostly for $\eta_2$ (0.87 against
0.50), and 0.04 [$-$0.07, 0.16] more than tuned GaSP's; they covered 77\% in the
20 replicates whose chains converged ($\hat R \le 1.01$). $\hat R$ exceeded 1.01
in 10, 1 and 3 replicates for \pkg{spDBL}, S-GaSP and GaSP; \pkg{spDBL}'s is
the worst of 20 conditional fits. A calibration took 9.4, 16.8 and 10.3 CPU
hours and gave 1.71, 0.024 and 0.032 bulk effective samples of its slowest
parameter per CPU second (means over replicates), so \pkg{spDBL} gave 72 and 54
times as many. With \pkg{RobustCalibration}'s default proposals, its chains accepted 11\% of moves, $\hat R$ exceeded
1.01 in 29 and 30 replicates and 1.1 in 14 and 21, and \pkg{spDBL} gave about
1{,}500 and 2{,}000 times more effective samples per CPU second, at similar
accuracy and coverage (64\% and 74\%).

With the 40 runs shipped with the package (Table~\ref{tab:e3-40}), every
method under-covered, \pkg{spDBL}'s most (54\% against 60\% and 67\%), at
relative RMSE 0.100 against 0.107 and 0.118. Only the $\eta_1$ differences
exceed sampling error: \pkg{spDBL} is less accurate than the GaSP variant and
covers less than both. \pkg{spDBL}'s intervals covered 45\% in the 17
replicates whose chains converged. The under-coverage does not come from the
emulator understating its error at the truth: at each replicate's true
parameters, the calibration emulator's pointwise 95\% predictive intervals
covered 98.8\% of the simulator output, and for each emulator draw the errors
whitened by the covariance the likelihood uses had mean square 0.40 (1 under
the model) and lag\nobreakdash-1 correlation over time 0.05
(\code{benchmarks/e3\_emulator\_error.R}). Ablations on replicates 1--10
(Table~\ref{tab:ablation-e3}) locate the cause. Jointly sampled coefficient
paths and 60 instead of 20 emulator draws changed nothing. With field data
whose simulator output was replaced by a draw from the emulator's own model,
so that \pkg{spDBL} is exactly specified, the intervals covered 93\% and the
relative RMSE halved: the inference is sound, and the narrow intervals come
from the emulator's systematic error, which shifts the posterior in ways its
predictive variance does not describe. Without the design-GP uncertainty the
intervals were 64\% wider and covered 77\% at the same accuracy; propagating
it, which corrects prediction intervals (E2), narrows the calibration
posterior here.

More training runs reduce the under-coverage (Table~\ref{tab:ablation-e3}). On
replicates 1--10, new Latin-hypercube designs of 40, 80 and 160 runs covered
77\%, 83\% and 93\% (the bundled 40 runs: 57\%; the 100-run design of
Table~\ref{tab:e3}: 83\%), with posterior means a median
0.96, 0.78 and 0.68 posterior standard deviations from the truth (1.42; 0.67 for
a calibrated posterior); a calibration took 12.7 and 17.2 CPU hours with 80 and
160 runs, against 8.9 with the bundled runs on these replicates. Adding 40 runs
where a short first-stage calibration put the posterior covered all 30
parameters, with the lowest error of any design (relative RMSE 0.052 against
0.086), at 12.0 CPU hours including the 1.3-hour first stage. The bundled design is also unfavourable: a fresh 40-run design covered
77\%. \citet{Banerjee2025} report 59 minutes for 30{,}000 draws of
their research code on an Apple M3 Max, against 26 minutes on 20 cores (7.1 CPU
hours) for Listing~\ref{lst:calibration}, which generates 80{,}000 post-warm-up NUTS draws
over 20 emulator draws and keeps 4{,}000.

\input{tables/appendix-tables}
\clearpage

\section{Spatial Prediction of Field Observations}\label{app:prediction}

Whereas \citet[Equations~10--12]{Banerjee2025} derive predictive distributions only for the emulated simulator output at untried mechanistic inputs over the fixed training grid $\mathcal{S} = \{\bm s_1, \ldots, \bm s_S\}$, their calibration development \citep[Equations~27--29]{Banerjee2025} addresses parameter estimation and smoothing exclusively at the $S_o$ observed field sites $\mathcal{S}_o = \{\bm s_{o,1}, \ldots, \bm s_{o,S_o}\} \subseteq \mathcal{S}$ (denoted $\tilde{\mathcal{S}}$ there). Predicting field measurements $z_t(\bm s_0)$ at an arbitrary spatial location $\bm s_0 \notin \mathcal{S}_o$ was not addressed in \citet{Banerjee2025} and is derived here by augmenting their joint posterior with latent states at $\bm s_0$ and applying conditional composition sampling.

Let $\mathcal{D} = \{\bm z_{1:T}, \bm Y_{1:T}\}$ denote the collection of $S_o \times 1$ field observation vectors $\bm z_{1:T} = \{\bm z_1, \ldots, \bm z_T\}$ at $\mathcal{S}_o$ and $N \times S$ simulator training outputs $\bm Y_{1:T} = \{\bm Y_1, \ldots, \bm Y_T\}$ over design inputs $\mathcal{X} = \{\bm x_1, \ldots, \bm x_N\}$ and grid $\mathcal{S}$. Let $\bm\Omega = \{\bm\eta, \bm\tau^2, \bm u_{0:T}, \bm y_{1:T}(\bm\eta), \bm\Theta, \bm\Sigma\}$ collect all calibration parameters and latent states at $\mathcal{S}_o$ from the joint posterior in \citet[Equation~28]{Banerjee2025}---where $\bm\tau^2 = (\tau_0^2, \tau_1^2, \ldots, \tau_T^2)^\top$, $\bm u_{0:T} = \{\bm u_0, \ldots, \bm u_T\}$ are the $S_o \times 1$ discrepancy vectors on $\mathcal{S}_o$, $\bm y_{1:T}(\bm\eta) = \{\bm y_1(\bm\eta), \ldots, \bm y_T(\bm\eta)\}$ are the $S_o \times 1$ latent emulator vectors on $\mathcal{S}_o$, and $\{\bm\Theta, \bm\Sigma\}$ are the emulator states and spatial covariance from~\eqref{eq:emulator}. When \code{calibration\_fit()} marginalises $\bm y_{1:T}(\bm\eta)$ and $\bm u_{0:T}$ in Stan to sample $(\bm\eta, \bm\tau^2)$ via NUTS for each of the $M$ emulator draws, draws of $\bm u_{0:T}$ and $\bm y_{1:T}(\bm\eta)$ on $\mathcal{S}_o$ are first recovered from their exact closed-form full conditionals \citep[Algorithms~7 and~8]{Banerjee2025} to form posterior draws of $\bm\Omega$.

To predict at a target location $\bm s_0 \notin \mathcal{S}_o$ and epoch $t \in \{1, \ldots, T\}$, let $\mathcal{S}_o^+ = \mathcal{S}_o \cup \{\bm s_0\}$ denote the augmented set of $S_o + 1$ sites, and define the augmented latent vectors $\bm u_k^+ = (\bm u_k^\top, u_k(\bm s_0))^\top \in \mathbf{R}^{S_o+1}$ ($k = 0, \ldots, T$) and $\bm y_k^+(\bm\eta) = (\bm y_k(\bm\eta)^\top, y_k(\bm\eta, \bm s_0))^\top \in \mathbf{R}^{S_o+1}$ ($k = 1, \ldots, T$). Generalising \citet[Equation~28]{Banerjee2025} from $\mathcal{S}_o$ to $\mathcal{S}_o^+$, the augmented joint posterior predictive density of $\bm\Omega^+ = \bm\Omega \cup \left\{\bm u_{0:T}(\bm s_0), \bm y_{1:T}(\bm\eta, \bm s_0), z_t(\bm s_0)\right\}$ given $\mathcal{D}$ is proportional to
\begin{equation}\label{eq:aug-post}
\begin{split}
p\left(\bm\Omega^+ \mid \mathcal{D}\right)
&\propto p(\bm\eta)\, p(\bm\Theta, \bm\Sigma \mid \bm Y_{1:T}) \times
  \mathcal{NG}\left(\bm u_0^+, \tau_0^{-2} \mid \bm m_0^{z,+}, \bm M_0^{z,+}, n_0^z, d_0^z\right)\\
&\qquad \times \prod_{k=1}^T \mathcal{NG}\left(\bm u_k^+, \tau_k^{-2} \mid \bm u_{k-1}^+, \bm U^+, b^k n_0^z, b^k d_0^z\right)\\
&\qquad\quad \times \prod_{k=1}^T \Big\{
  \mathcal{N}_{S_o+1}\left(\bm y_k^+(\bm\eta) \mid \tilde{\bm\mu}_k^+(\bm\eta), \tilde{\bm\Sigma}_k^+(\bm\eta)\right)\\
&\qquad\qquad \times \prod_{i=1}^{S_o} \mathcal{N}\left(z_k(\bm s_{o,i}) \mid y_k(\bm\eta, \bm s_{o,i}) + u_k(\bm s_{o,i}), \tau_k^2\right)
  \Big\}\\
&\qquad\quad\quad \times \mathcal{N}\left(z_t(\bm s_0) \mid y_t(\bm\eta, \bm s_0) + u_t(\bm s_0), \tau_t^2\right),
\end{split}
\end{equation}
where $\mathcal{NG}(\cdot)$ denotes the Normal--Gamma density, $\bm U$ is user-supplied (so $p(\bm\eta, \bm\rho)$ in \citet[Equation~28]{Banerjee2025} reduces to $p(\bm\eta)$), $\bm m_0^{z,+} = \left((\bm m_0^z)^\top, m_0^z(\bm s_0)\right)^\top$,
\[
\bm M_0^{z,+} = \begin{bmatrix} \bm M_0^z & \bm m_{0,o}^z(\bm s_0) \\ \bm m_{0,o}^z(\bm s_0)^\top & m_{00}^z(\bm s_0) \end{bmatrix}, \qquad
\bm U^+ = \begin{bmatrix} \bm U & \bm c_0(\bm s_0) \\ \bm c_0(\bm s_0)^\top & 1 \end{bmatrix},
\]
with $\bm m_{0,o}^z(\bm s_0) = \bm M_0^z(\mathcal{S}_o, \bm s_0)$, $m_{00}^z(\bm s_0) = M_0^z(\bm s_0, \bm s_0)$, $\bm c_0(\bm s_0) = [\mathcal{C}(\bm s_{o,1}, \bm s_0), \ldots, \mathcal{C}(\bm s_{o,S_o}, \bm s_0)]^\top$, $\mathcal{C}(\bm s_0, \bm s_0) = 1$, and $\tilde{\bm\mu}_k^+(\bm\eta), \tilde{\bm\Sigma}_k^+(\bm\eta)$ denoting the $(S_o+1)$-dimensional emulator predictive mean and covariance on $\mathcal{S}_o^+$ (with the off-grid case $\bm s_0 \notin \mathcal{S}$ also conditioning on the imputed state column $\bm\theta_k(\bm s_0)$).

Integrating~\eqref{eq:aug-post} over $\bm u_{0:T}(\bm s_0)$, $\bm y_{1:T}(\bm\eta, \bm s_0)$, and $z_t(\bm s_0)$ recovers the calibration posterior $p(\bm\Omega \mid \mathcal{D})$ of \citet[Equation~28]{Banerjee2025}. Conversely, dividing~\eqref{eq:aug-post} by $p(\bm\Omega \mid \mathcal{D})$ and marginalising out $y_k(\bm\eta, \bm s_0)$ for $k \neq t$ and $u_k(\bm s_0)$ for $k > t$ yields the joint posterior predictive decomposition by conditional composition:
\begin{equation}\label{eq:comp-decomp}
\begin{split}
p\left(z_t(\bm s_0), y_t(\bm\eta, \bm s_0), u_t(\bm s_0) \mid \mathcal{D}\right)
&= \int p\left(z_t(\bm s_0) \mid y_t(\bm\eta, \bm s_0), u_t(\bm s_0),
  \mathcal{D}, \bm\Omega\right)\\
&\qquad\times p\left(y_t(\bm\eta, \bm s_0), u_t(\bm s_0) \mid \mathcal{D},
  \bm\Omega\right)\, p(\bm\Omega \mid \mathcal{D})\, d\bm\Omega.
\end{split}
\end{equation}
Because the emulator terms $\bm y_{1:T}^+(\bm\eta)$ and discrepancy terms $\bm u_{0:T}^+$ appear in disjoint product factors in~\eqref{eq:aug-post} and $z_t(\bm s_0)$ is unobserved, $y_t(\bm\eta, \bm s_0)$ and $u_t(\bm s_0)$ are conditionally independent given $(\mathcal{D}, \bm\Omega)$:
\begin{equation}\label{eq:cond-indep}
\begin{split}
p\left(y_t(\bm\eta, \bm s_0), u_t(\bm s_0) \mid \mathcal{D}, \bm\Omega\right)
&= p\left(y_t(\bm\eta, \bm s_0) \mid \bm y_t(\bm\eta), \bm Y_t, \bm\Theta_t,
  \bm\Sigma, \bm\eta\right) \times p\left(u_t(\bm s_0) \mid \bm u_{0:T}, \bm\tau^2\right).
\end{split}
\end{equation}

\subsection{Conditional distribution of \texorpdfstring{$y_t(\bm \eta, \bm s_0)$}{yt(eta, s0)}.}
Under~\eqref{eq:emulator} and \citet[Equations~10--11]{Banerjee2025}, the emulated field at time $t$ conditional on $(\bm\Theta_t, \bm\Sigma, \bm\eta)$ has mean $\bm f_t(\bm x, \bm s)^\top \bm\theta_t(\bm s)$ (with $\bm\theta_t(\bm s) \in \mathbf{R}^p$ the state column at site $\bm s$) and separable covariance $\operatorname{Cov}\left(y_t(\bm x, \bm s), y_t(\bm x', \bm s') \mid \bm\Theta_t, \bm\Sigma\right) = \mathcal{C}_{\mathcal{X}}(\bm x, \bm x')\,\Sigma(\bm s, \bm s')$, where $\mathcal{C}_{\mathcal{X}}(\cdot,\cdot)$ is the input-space correlation function with $\mathcal{C}_{\mathcal{X}}(\bm x, \bm x) = 1$, $\bm V = [\mathcal{C}_{\mathcal{X}}(\bm x_i, \bm x_j)]_{i,j=1}^N$, and $\bm J_t(\bm\eta) = [\mathcal{C}_{\mathcal{X}}(\bm x_1, \bm\eta), \ldots, \mathcal{C}_{\mathcal{X}}(\bm x_N, \bm\eta)]^\top \in \mathbf{R}^N$. Define the input-space kriging weight vector $\bm\kappa_t(\bm\eta) = \bm V^{-1} \bm J_t(\bm\eta) \in \mathbf{R}^N$, Schur complement $c_t(\bm\eta) = 1 - \bm J_t(\bm\eta)^\top \bm\kappa_t(\bm\eta) \in [0, 1]$ (with $c_t(\bm\eta) > 0$ for $\bm\eta \notin \mathcal{X}$), and $q_t(\bm\eta) = 1 - c_t(\bm\eta)$. At the observed sites $\mathcal{S}_o$, $\bm y_t(\bm\eta) \mid \bm Y_t, \bm\Theta_t, \bm\Sigma, \bm\eta \sim \mathcal{N}_{S_o}\left(\tilde{\bm\mu}_t(\bm\eta), c_t(\bm\eta)\bm\Sigma_o\right)$, where $\bm\Sigma_o = \bm\Sigma(\mathcal{S}_o)$ is the $S_o \times S_o$ submatrix of $\bm\Sigma$ on $\mathcal{S}_o$ and
\[
\tilde{\bm\mu}_t(\bm\eta) = \bm\Theta_t(\mathcal{S}_o)^\top
  \tilde{\bm F}_t(\bm\eta)^\top + \left(\bm Y_t(\mathcal{S}_o) - \bm F_t
  \bm\Theta_t(\mathcal{S}_o)\right)^\top \bm\kappa_t(\bm\eta),
\]
with $\bm\Theta_t(\mathcal{S}_o)$ ($p \times S_o$) and $\bm Y_t(\mathcal{S}_o)$ ($N \times S_o$) selecting the columns indexed by $\mathcal{S}_o$, and $\tilde{\bm F}_t(\bm\eta)$ ($1 \times p$) denoting the design row at $\bm\eta$ (including the design-GP draw when used). We consider two cases for $\bm s_0$:
{\footnotesize
\begin{algorithm}[p]
\caption{Conditional Spatial Sampling of the Emulated Field $y_t(\bm\eta, \bm s_0)$}
\label{alg:y-pred}
\begin{algorithmic}[1]
\Require Target site $\bm s_0 \notin \mathcal{S}_o$, target epoch set $\mathcal{T} = \{t\}$ (with $t_{\max} = t$) or full horizon $\mathcal{T} = \{1, \ldots, T\}$ (with $t_{\max} = T$), training data $\{\bm Y_{1:T}, \bm F_{1:T}\}$, and posterior draws $\left\{\bm\eta^{(l)}, \bm y_{1:T}^{(l)}(\bm\eta^{(l)}), \bm\Theta_{0:T}^{(l)}, \bm\Sigma^{(l)}\right\}_{l=1}^L$ with design rows $\tilde{\bm F}_{1:T}(\bm\eta^{(l)})$.
\Ensure Conditional posterior predictive draws $\left\{y_k^{(l)}(\bm\eta^{(l)}, \bm s_0) : k \in \mathcal{T}\right\}_{l=1}^L$.
\If{$\bm s_0 \notin \mathcal{S}$} \Comment{Precompute grid-level spatial kriging weights for off-grid GP emulator}
  \State Compute $\bm\lambda(\bm s_0) \gets \bm R^{-1}\bm r(\bm s_0)$, $v_{\mathcal{S}}(\bm s_0) \gets 1 - \bm r(\bm s_0)^\top\bm\lambda(\bm s_0)$, and $\bm\lambda_o(\bm s_0) \gets \bm R_o^{-1}\bm r_o(\bm s_0)$.
\EndIf
\For{$l = 1$ to $L$}
  \If{$\bm s_0 \notin \mathcal{S}$} \Comment{Impute latent state trajectory at $\bm s_0$ up to $t_{\max}$}
    \State Draw $\bm\theta_0^{(l)}(\bm s_0) \sim \mathcal{N}_p\left(\bm m_0(\bm s_0) + (\bm\Theta_0^{(l)} - \bm m_0)\bm\lambda(\bm s_0),\; \sigma^{2,(l)} v_{\mathcal{S}}(\bm s_0)\bm M_0\right)$.
    \For{$k = 1$ to $t_{\max}$}
      \State Draw $\bm\theta_k^{(l)}(\bm s_0) \sim \mathcal{N}_p\left(\bm G\bm\theta_{k-1}^{(l)}(\bm s_0) + (\bm\Theta_k^{(l)} - \bm G\bm\Theta_{k-1}^{(l)})\bm\lambda(\bm s_0),\; \sigma^{2,(l)} v_{\mathcal{S}}(\bm s_0)\bm W\right)$.
    \EndFor
  \EndIf
  \For{$k \in \mathcal{T}$}
    \State Compute $\bm\kappa_k^{(l)} \gets \bm V_k^{-1}\bm J_k(\bm\eta^{(l)})$, $c_k^{(l)} \gets 1 - \bm J_k(\bm\eta^{(l)})^\top\bm\kappa_k^{(l)}$, and $q_k^{(l)} \gets 1 - c_k^{(l)}$.
    \State Compute $\tilde{\bm\mu}_k^{(l)} \gets \bm\Theta_k^{(l)}(\mathcal{S}_o)^\top\tilde{\bm F}_k(\bm\eta^{(l)})^\top + \left(\bm Y_k(\mathcal{S}_o) - \bm F_k\bm\Theta_k^{(l)}(\mathcal{S}_o)\right)^\top\bm\kappa_k^{(l)}$.
    \If{$\bm s_0 \in \mathcal{S} \setminus \mathcal{S}_o$} \Comment{Case 1: on-grid prediction via~\eqref{eq:y-ongrid}}
      \State Compute $\bm\gamma_y^{(l)}(\bm s_0) \gets \left(\bm\Sigma_o^{(l)}\right)^{-1}\bm\sigma_o^{(l)}(\bm s_0)$ and $v_y^{(l)}(\bm s_0) \gets \sigma_{00}^{(l)}(\bm s_0) - \bm\sigma_o^{(l)}(\bm s_0)^\top\bm\gamma_y^{(l)}(\bm s_0)$.
      \State $\tilde{\mu}_k^{(l)}(\bm s_0) \gets \tilde{\bm f}_k(\bm\eta^{(l)}, \bm s_0)^\top\bm\theta_k^{(l)}(\bm s_0) + \left(\bm\kappa_k^{(l)}\right)^\top\left(\bm Y_k(\cdot, \bm s_0) - \bm F_k\bm\theta_k^{(l)}(\bm s_0)\right)$.
      \State $\mu_{y,k}^{(l)}(\bm s_0) \gets \tilde{\mu}_k^{(l)}(\bm s_0) + \bm\gamma_y^{(l)}(\bm s_0)^\top\left(\bm y_k^{(l)}(\bm\eta^{(l)}) - \tilde{\bm\mu}_k^{(l)}\right)$, \quad $\sigma_{y,k}^{2,(l)}(\bm s_0) \gets c_k^{(l)}\, v_y^{(l)}(\bm s_0)$.
    \Else \Comment{Case 2: off-grid prediction ($\bm s_0 \notin \mathcal{S}$, $\bm\Sigma^{(l)} = \sigma^{2,(l)}\bm R$) via~\eqref{eq:y-offgrid}}
      \State $\tilde{\mu}_k^{\mathrm{krig},(l)}(\bm s_0) \gets \tilde{\bm f}_k(\bm\eta^{(l)}, \bm s_0)^\top\bm\theta_k^{(l)}(\bm s_0) + \left(\bm\kappa_k^{(l)}\right)^\top\left(\bm Y_k - \bm F_k\bm\Theta_k^{(l)}\right)\bm\lambda(\bm s_0)$.
      \State $v_{k,0}^{(l)}(\bm s_0) \gets 1 - q_k^{(l)}\,\bm r(\bm s_0)^\top\bm\lambda(\bm s_0)$.
      \State $\mu_{y,k}^{(l)}(\bm s_0) \gets \tilde{\mu}_k^{\mathrm{krig},(l)}(\bm s_0) + \bm\lambda_o(\bm s_0)^\top\left(\bm y_k^{(l)}(\bm\eta^{(l)}) - \tilde{\bm\mu}_k^{(l)}\right)$.
      \State $\sigma_{y,k}^{2,(l)}(\bm s_0) \gets \sigma^{2,(l)}\left(v_{k,0}^{(l)}(\bm s_0) - c_k^{(l)}\,\bm r_o(\bm s_0)^\top\bm\lambda_o(\bm s_0)\right)$.
    \EndIf
    \State Draw $y_k^{(l)}(\bm\eta^{(l)}, \bm s_0) \sim \mathcal{N}\left(\mu_{y,k}^{(l)}(\bm s_0),\, \sigma_{y,k}^{2,(l)}(\bm s_0)\right)$.
  \EndFor
\EndFor
\end{algorithmic}
\end{algorithm}
}

\subsection{Case 1: \texorpdfstring{On-grid prediction ($\bm s_0 \in \mathcal{S} \setminus \mathcal{S}_o$; both $\bm\Sigma \sim \mathcal{IW}(n_0, \bm D_0)$ and $\bm\Sigma = \sigma^2 \bm R$).}{On-grid prediction}}
Here the training column $\bm Y_t(\cdot, \bm s_0) \in \mathbf{R}^N$ and state column $\bm\theta_t(\bm s_0) \in \mathbf{R}^p$ are available on the emulator grid. Applying \citet[Equation~11]{Banerjee2025} jointly on $\mathcal{S}_o^+ = \mathcal{S}_o \cup \{\bm s_0\}$ specifies $\tilde{\bm\mu}_t^+(\bm\eta)$ and $\tilde{\bm\Sigma}_t^+(\bm\eta)$ in~\eqref{eq:aug-post} as
\[
\bm y_t^+(\bm\eta) = \begin{bmatrix} \bm y_t(\bm\eta) \\ y_t(\bm\eta, \bm s_0) \end{bmatrix}
\,\Bigg|\, \bm Y_t, \bm\Theta_t, \bm\Sigma, \bm\eta
\sim \mathcal{N}_{S_o+1}\!\left(
\begin{bmatrix} \tilde{\bm\mu}_t(\bm\eta) \\ \tilde{\mu}_t(\bm\eta, \bm s_0) \end{bmatrix},\;
c_t(\bm\eta)\begin{bmatrix}
\bm\Sigma_o & \bm\sigma_o(\bm s_0) \\
\bm\sigma_o(\bm s_0)^\top & \sigma_{00}(\bm s_0)
\end{bmatrix}\right),
\]
where $\bm\sigma_o(\bm s_0) = \bm\Sigma(\mathcal{S}_o, \bm s_0) \in \mathbf{R}^{S_o}$, $\sigma_{00}(\bm s_0) = \Sigma(\bm s_0, \bm s_0)$, and $\tilde{\mu}_t(\bm\eta, \bm s_0) = \tilde{\bm f}_t(\bm\eta, \bm s_0)^\top \bm\theta_t(\bm s_0) + \bm\kappa_t(\bm\eta)^\top \left(\bm Y_t(\cdot, \bm s_0) - \bm F_t\bm\theta_t(\bm s_0)\right)$. Conditioning further on $\bm y_t(\bm\eta)$, the scalar $c_t(\bm\eta)$ cancels in the kriging weights and yields
\begin{equation}\label{eq:y-ongrid}
y_t(\bm\eta, \bm s_0) \mid \bm y_t(\bm\eta), \bm Y_t, \bm\Theta_t, \bm\Sigma, \bm\eta
\sim \mathcal{N}\left(\mu_{y,t}(\bm s_0),\, \sigma_{y,t}^2(\bm s_0)\right),
\end{equation}
where the mean is $\mu_{y,t}(\bm s_0) = \tilde{\mu}_t(\bm\eta, \bm s_0) + \bm\sigma_o(\bm s_0)^\top \bm\Sigma_o^{-1} \left(\bm y_t(\bm\eta) - \tilde{\bm\mu}_t(\bm\eta)\right)$ and the variance is $\sigma_{y,t}^2(\bm s_0) = c_t(\bm\eta)\left(\sigma_{00}(\bm s_0) - \bm\sigma_o(\bm s_0)^\top \bm\Sigma_o^{-1}\bm\sigma_o(\bm s_0)\right)$.

\subsection{Case 2: Off-grid prediction (\texorpdfstring{$\bm s_0 \notin \mathcal{S}$; spatial GP emulator $\bm\Sigma = \sigma^2 \bm R$}{Off-grid prediction}).}
When $\bm s_0 \notin \mathcal{S}$, spatial interpolation requires the structured spatial covariance $\bm\Sigma = \sigma^2 \bm R$ induced by a spatial correlation function $\rho(\bm s, \bm s')$ with $\rho(\bm s, \bm s) = 1$ \citep[Section~2.3]{Banerjee2025}. Let $\bm r(\bm s_0) = [\rho(\bm s_1, \bm s_0), \ldots, \rho(\bm s_S, \bm s_0)]^\top$ and $\bm r_o(\bm s_0) = [\rho(\bm s_{o,1}, \bm s_0), \ldots, \rho(\bm s_{o,S_o}, \bm s_0)]^\top$, and define the spatial kriging weight vector $\bm\lambda(\bm s_0) = \bm R^{-1}\bm r(\bm s_0) \in \mathbf{R}^S$ and Schur complement $v_{\mathcal{S}}(\bm s_0) = 1 - \bm r(\bm s_0)^\top \bm\lambda(\bm s_0)$. Because no simulator run exists at $\bm s_0 \notin \mathcal{S}$, the latent state column $\bm\theta_t(\bm s_0)$ is first imputed by spatial kriging across the columns of $\bm\Theta_{0:t}$ in~\eqref{eq:emulator} sequentially for $k = 0, 1, \ldots, t$:
\begin{align*}
\bm\theta_0(\bm s_0) \mid \bm\Theta_0, \sigma^2
  &\sim \mathcal{N}_p\!\left(\bm m_0(\bm s_0) + (\bm\Theta_0 - \bm m_0)\bm\lambda(\bm s_0),\;
    \sigma^2 v_{\mathcal{S}}(\bm s_0)\bm M_0\right),\\
\bm\theta_k(\bm s_0) \mid \bm\theta_{k-1}(\bm s_0), \bm\Theta_{k-1:k}, \sigma^2
  &\sim \mathcal{N}_p\!\left(\bm G\bm\theta_{k-1}(\bm s_0)
    + (\bm\Theta_k - \bm G\bm\Theta_{k-1})\bm\lambda(\bm s_0),\;
    \sigma^2 v_{\mathcal{S}}(\bm s_0)\bm W\right).
\end{align*}
Next, under the separable residual process
\[
\operatorname{Cov}\left(\mathcal{E}_t(\bm x, \bm s), \mathcal{E}_t(\bm x', \bm s') \mid \sigma^2\right) = \sigma^2 \mathcal{C}_{\mathcal{X}}(\bm x, \bm x')\rho(\bm s, \bm s'),
\]
conditioning $\bm y_t^+(\bm\eta)$ first on $\bm Y_t$ (using $\mathcal{S}_o \subseteq \mathcal{S}$, so $\bm r(\bm s_0)$ restricted to $\mathcal{S}_o$ equals $\bm r_o(\bm s_0)$) specifies $\tilde{\bm\mu}_t^+(\bm\eta)$ and $\tilde{\bm\Sigma}_t^+(\bm\eta)$ in~\eqref{eq:aug-post} as
\begin{align*}
\bm y_t^+(\bm\eta) = \begin{bmatrix} \bm y_t(\bm\eta) \\ y_t(\bm\eta, \bm s_0) \end{bmatrix}
\,\Bigg|\, \bm Y_t, \bm\Theta_t, \bm\theta_t(\bm s_0), \sigma^2, \bm\eta
&\sim \mathcal{N}_{S_o+1}\!\Bigg(
\begin{bmatrix} \tilde{\bm\mu}_t(\bm\eta) \\
  \tilde{\mu}_t^{\mathrm{krig}}(\bm\eta, \bm s_0) \end{bmatrix},\\
&\qquad\quad \sigma^2\begin{bmatrix}
c_t(\bm\eta)\bm R_o & c_t(\bm\eta)\bm r_o(\bm s_0) \\
c_t(\bm\eta)\bm r_o(\bm s_0)^\top & v_{t,0}(\bm s_0)
\end{bmatrix}\Bigg),
\end{align*}
where $\bm R_o = \bm R(\mathcal{S}_o)$, $v_{t,0}(\bm s_0) = 1 - q_t(\bm\eta) \bm r(\bm s_0)^\top\bm\lambda(\bm s_0)$, and $\tilde{\mu}_t^{\mathrm{krig}}(\bm\eta, \bm s_0) = \tilde{\bm f}_t(\bm\eta, \bm s_0)^\top\bm\theta_t(\bm s_0) + \bm\kappa_t(\bm\eta)^\top(\bm Y_t - \bm F_t\bm\Theta_t)\bm\lambda(\bm s_0)$. Conditioning further on $\bm y_t(\bm\eta)$ gives
\begin{equation}\label{eq:y-offgrid}
y_t(\bm\eta, \bm s_0) \mid \bm y_t(\bm\eta), \bm Y_t, \bm\Theta_t,
\bm\theta_t(\bm s_0), \sigma^2, \bm\eta
\sim \mathcal{N}\left(\mu_{y,t}(\bm s_0),\, \sigma_{y,t}^2(\bm s_0)\right),
\end{equation}
where the mean at $\bm s_0$ is $\mu_{y,t}(\bm s_0) = \tilde{\mu}_t^{\mathrm{krig}}(\bm\eta, \bm s_0) + \bm r_o(\bm s_0)^\top \bm R_o^{-1} \left(\bm y_t(\bm\eta) - \tilde{\bm\mu}_t(\bm\eta)\right)$ and the variance is $\sigma_{y,t}^2(\bm s_0) = \sigma^2\left(v_{t,0}(\bm s_0) - c_t(\bm\eta)\,\bm r_o(\bm s_0)^\top \bm R_o^{-1}\bm r_o(\bm s_0)\right)$. Algorithm \ref{alg:y-pred} describes the conditional spatial sampling of the emulated field $y_t(\bm\eta, \bm s_0)$.

\subsection{Conditional distribution of \texorpdfstring{$u_t(\bm s_0)$}{ut(s0)}.}
From~\eqref{eq:calibration} and the Normal--Gamma factors in~\eqref{eq:aug-post}, the augmented spatial discrepancy evolves as a Gaussian-process random walk $\bm u_0^+ \mid \tau_0^2 \sim \mathcal{N}_{S_o+1}\left(\bm m_0^{z,+},\, \tau_0^2 \bm M_0^{z,+}\right)$ with independent increments $\bm\varepsilon_k^{u,+} = \bm u_k^+ - \bm u_{k-1}^+ \mid \tau_k^2 \stackrel{\text{ind.}}{\sim} \mathcal{N}_{S_o+1}\left(\bm 0,\, \tau_k^2 \bm U^+\right)$ for $k = 1, \ldots, T$. Conditioning on $\bm u_{0:T}$ and $\bm\tau^2$, partitioning $\bm u_0^+$ gives $u_0(\bm s_0) \mid \bm u_{0:T}, \bm\tau^2 \sim \mathcal{N}\left(\mu_{u,0}(\bm s_0), \sigma_{u,0}^2(\bm s_0)\right)$ with mean $\mu_{u,0}(\bm s_0) = m_0^z(\bm s_0) + \bm m_{0,o}^z(\bm s_0)^\top (\bm M_0^z)^{-1}(\bm u_0 - \bm m_0^z)$ and variance $\sigma_{u,0}^2(\bm s_0) = \tau_0^2\left(m_{00}^z(\bm s_0) - \bm m_{0,o}^z(\bm s_0)^\top (\bm M_0^z)^{-1}\bm m_{0,o}^z(\bm s_0)\right)$ (where $\mu_{u,0}(\bm s_0) = \sigma_{u,0}^2(\bm s_0) = 0$ if $u_0(\cdot) \equiv 0$ is fixed), while partitioning $\bm\varepsilon_k^{u,+} = \left((\bm\varepsilon_k^u)^\top, \varepsilon_k^u(\bm s_0)\right)^\top$ implies
\[
\varepsilon_k^u(\bm s_0) \mid \bm u_{0:T}, \bm\tau^2 \stackrel{\text{ind.}}{\sim}
\mathcal{N}\left(\bm\gamma_u(\bm s_0)^\top(\bm u_k - \bm u_{k-1}),\;
\tau_k^2 v_u(\bm s_0)\right),
\]
where $\bm\gamma_u(\bm s_0) = \bm U^{-1}\bm c_0(\bm s_0) \in \mathbf{R}^{S_o}$ and $v_u(\bm s_0) = 1 - \bm c_0(\bm s_0)^\top\bm\gamma_u(\bm s_0)$. Writing $u_t(\bm s_0) = u_0(\bm s_0) + \sum_{k=1}^t \varepsilon_k^u(\bm s_0)$ and summing the conditionally independent normal terms, the kriging increments telescope across $k = 1, \ldots, t$ while the innovation variances accumulate:
\begin{equation}\label{eq:u-cond}
u_t(\bm s_0) \mid \bm u_{0:T}, \bm\tau^2 \sim \mathcal{N}\left(\mu_{u,t}(\bm s_0),\,
\sigma_{u,t}^2(\bm s_0)\right),
\end{equation}
where the mean is $\mu_{u,t}(\bm s_0) = \mu_{u,0}(\bm s_0) + \bm\gamma_u(\bm s_0)^\top(\bm u_t - \bm u_0)$ and variance $\sigma_{u,t}^2(\bm s_0) = \sigma_{u,0}^2(\bm s_0) + v_u(\bm s_0)\sum_{k=1}^t \tau_k^2$. (If joint trajectories $u_{0:T}(\bm s_0)$ across all $t = 0, \ldots, T$ are required rather than a single epoch $t$, one draws $u_0(\bm s_0)$ first and accumulates $u_k(\bm s_0) = u_{k-1}(\bm s_0) + \varepsilon_k^u(\bm s_0)$ sequentially for $k = 1, \ldots, T$.) Algorithm~\ref{alg:u-pred} describes conditional spatial sampling of the discrepancy process $u_t(\bm s_0)$.
\begin{algorithm}[t]
\caption{Conditional Spatial Sampling of the Discrepancy Process $u_t(\bm s_0)$}
\label{alg:u-pred}
\begin{algorithmic}[1]
\Require Target site $\bm s_0 \notin \mathcal{S}_o$, target epoch set $\mathcal{T} = \{t\}$ or full horizon $\mathcal{T} = \{1, \ldots, T\}$, spatial correlation matrix $\bm U$ and cross-correlation vector $\bm c_0(\bm s_0)$, and posterior draws $\left\{\bm u_{0:T}^{(l)}, \bm\tau^{2,(l)}\right\}_{l=1}^L$.
\Ensure Conditional posterior predictive draws $\left\{u_k^{(l)}(\bm s_0) : k \in \mathcal{T}\right\}_{l=1}^L$.
\State Precompute $\bm\gamma_u(\bm s_0) \gets \bm U^{-1}\bm c_0(\bm s_0)$ and $v_u(\bm s_0) \gets 1 - \bm c_0(\bm s_0)^\top\bm\gamma_u(\bm s_0)$.
\If{$u_0(\cdot)$ is random}
  \State Precompute $\bm\gamma_{u,0}(\bm s_0) \gets (\bm M_0^z)^{-1}\bm m_{0,o}^z(\bm s_0)$ and $v_{u,0}(\bm s_0) \gets m_{00}^z(\bm s_0) - \bm m_{0,o}^z(\bm s_0)^\top\bm\gamma_{u,0}(\bm s_0)$.
\EndIf
\For{$l = 1$ to $L$}
  \If{$u_0(\cdot) \equiv 0$ is fixed}
    \State Set $\mu_{u,0}^{(l)}(\bm s_0) \gets 0$ and $\sigma_{u,0}^{2,(l)}(\bm s_0) \gets 0$.
  \Else
    \State $\mu_{u,0}^{(l)}(\bm s_0) \gets m_0^z(\bm s_0) + \bm\gamma_{u,0}(\bm s_0)^\top\left(\bm u_0^{(l)} - \bm m_0^z\right)$, \quad $\sigma_{u,0}^{2,(l)}(\bm s_0) \gets \tau_0^{2,(l)}\, v_{u,0}(\bm s_0)$.
  \EndIf
  \If{$\mathcal{T} = \{t\}$} \Comment{Single-epoch marginal draw via~\eqref{eq:u-cond}}
    \State $\mu_{u,t}^{(l)}(\bm s_0) \gets \mu_{u,0}^{(l)}(\bm s_0) + \bm\gamma_u(\bm s_0)^\top\left(\bm u_t^{(l)} - \bm u_0^{(l)}\right)$, \quad $\sigma_{u,t}^{2,(l)}(\bm s_0) \gets \sigma_{u,0}^{2,(l)}(\bm s_0) + v_u(\bm s_0)\sum_{k=1}^t \tau_k^{2,(l)}$.
    \State Draw $u_t^{(l)}(\bm s_0) \sim \mathcal{N}\left(\mu_{u,t}^{(l)}(\bm s_0),\, \sigma_{u,t}^{2,(l)}(\bm s_0)\right)$.
  \Else \Comment{Full-horizon trajectory across $\mathcal{T} = \{1, \ldots, T\}$}
    \State Draw $u_0^{(l)}(\bm s_0) \sim \mathcal{N}\left(\mu_{u,0}^{(l)}(\bm s_0),\, \sigma_{u,0}^{2,(l)}(\bm s_0)\right)$.
    \For{$k = 1$ to $T$}
      \State Draw $\varepsilon_k^{u,(l)}(\bm s_0) \sim \mathcal{N}\left(\bm\gamma_u(\bm s_0)^\top\left(\bm u_k^{(l)} - \bm u_{k-1}^{(l)}\right),\, \tau_k^{2,(l)} v_u(\bm s_0)\right)$ and set $u_k^{(l)}(\bm s_0) \gets u_{k-1}^{(l)}(\bm s_0) + \varepsilon_k^{u,(l)}(\bm s_0)$.
    \EndFor
  \EndIf
\EndFor
\end{algorithmic}
\end{algorithm}

\subsection{Conditional distribution of \texorpdfstring{$z_t(\bm s_0)$}{zt(s0)} and composition sampler.}
From the final observation factor in~\eqref{eq:aug-post}, $z_t(\bm s_0)$ is conditionally independent of $\mathcal{D}$ and all remaining components of $\bm\Omega$ given $\left(\bm\eta, y_t(\bm\eta, \bm s_0), u_t(\bm s_0), \tau_t^2\right)$, so
\begin{equation}\label{eq:z-cond}
z_t(\bm s_0) \mid \bm\eta, y_t(\bm\eta, \bm s_0), u_t(\bm s_0), \tau_t^2
\sim \mathcal{N}\left(y_t(\bm\eta, \bm s_0) + u_t(\bm s_0),\, \tau_t^2\right).
\end{equation}
Evaluating~\eqref{eq:comp-decomp}--\eqref{eq:z-cond} across each posterior draw $\bm\Omega^{(l)}$ ($l = 1, \ldots, L$ pooled over the $M$ emulator draws) yields draws from $\left[z_t(\bm s_0), y_t(\bm\eta, \bm s_0), u_t(\bm s_0) \mid \mathcal{D}\right]$ via three univariate normal draws, separated into Algorithms~\ref{alg:y-pred}--\ref{alg:z-pred}:
\begin{enumerate}
\item Draw $y_t^{(l)}(\bm\eta^{(l)}, \bm s_0) \sim \mathcal{N}\left(\mu_{y,t}^{(l)}(\bm s_0),\, \sigma_{y,t}^{2,(l)}(\bm s_0)\right)$ from~\eqref{eq:y-ongrid} or~\eqref{eq:y-offgrid} (Algorithm~\ref{alg:y-pred}).
\item Independently draw $u_t^{(l)}(\bm s_0) \sim \mathcal{N}\left(\mu_{u,t}^{(l)}(\bm s_0),\, \sigma_{u,t}^{2,(l)}(\bm s_0)\right)$ from~\eqref{eq:u-cond} (Algorithm~\ref{alg:u-pred}).
\item Matching index $l$ across Algorithms~\ref{alg:y-pred} and~\ref{alg:u-pred}, draw
\[
z_t^{(l)}(\bm s_0) \sim \mathcal{N}\left(y_t^{(l)}(\bm\eta^{(l)}, \bm s_0) + u_t^{(l)}(\bm s_0),\, \tau_t^{2,(l)}\right)
\]
from~\eqref{eq:z-cond} (Algorithm~\ref{alg:z-pred}).
\end{enumerate}

\begin{algorithm}[t]
\caption{Posterior Predictive Composition Sampler for Field Observations $z_t(\bm s_0)$}
\label{alg:z-pred}
\begin{algorithmic}[1]
\Require Paired conditional draws $\left\{\bm\eta^{(l)}, y_k^{(l)}(\bm\eta^{(l)}, \bm s_0) : k \in \mathcal{T}\right\}_{l=1}^L$\newline from Algorithm~\ref{alg:y-pred}, $\left\{u_k^{(l)}(\bm s_0) : k \in \mathcal{T}\right\}_{l=1}^L$ from Algorithm~\ref{alg:u-pred}, and posterior variance draws $\left\{\tau_k^{2,(l)} : k \in \mathcal{T}\right\}_{l=1}^L$.
\Ensure Posterior predictive draws $\left\{z_k^{(l)}(\bm s_0) : k \in \mathcal{T}\right\}_{l=1}^L$ from $p\left(z_k(\bm s_0) \mid \mathcal{D}\right)$.
\For{$l = 1$ to $L$}
  \For{$k \in \mathcal{T}$}
    \State Draw $z_k^{(l)}(\bm s_0) \sim \mathcal{N}\left(y_k^{(l)}(\bm\eta^{(l)}, \bm s_0) + u_k^{(l)}(\bm s_0),\, \tau_k^{2,(l)}\right)$ via~\eqref{eq:z-cond}.
  \EndFor
\EndFor
\end{algorithmic}
\end{algorithm}

\subsection{Prediction accuracy for field measurements.}\label{app:prediction-accuracy}
\begin{figure}[t]
\centering
\includegraphics[width=\linewidth]{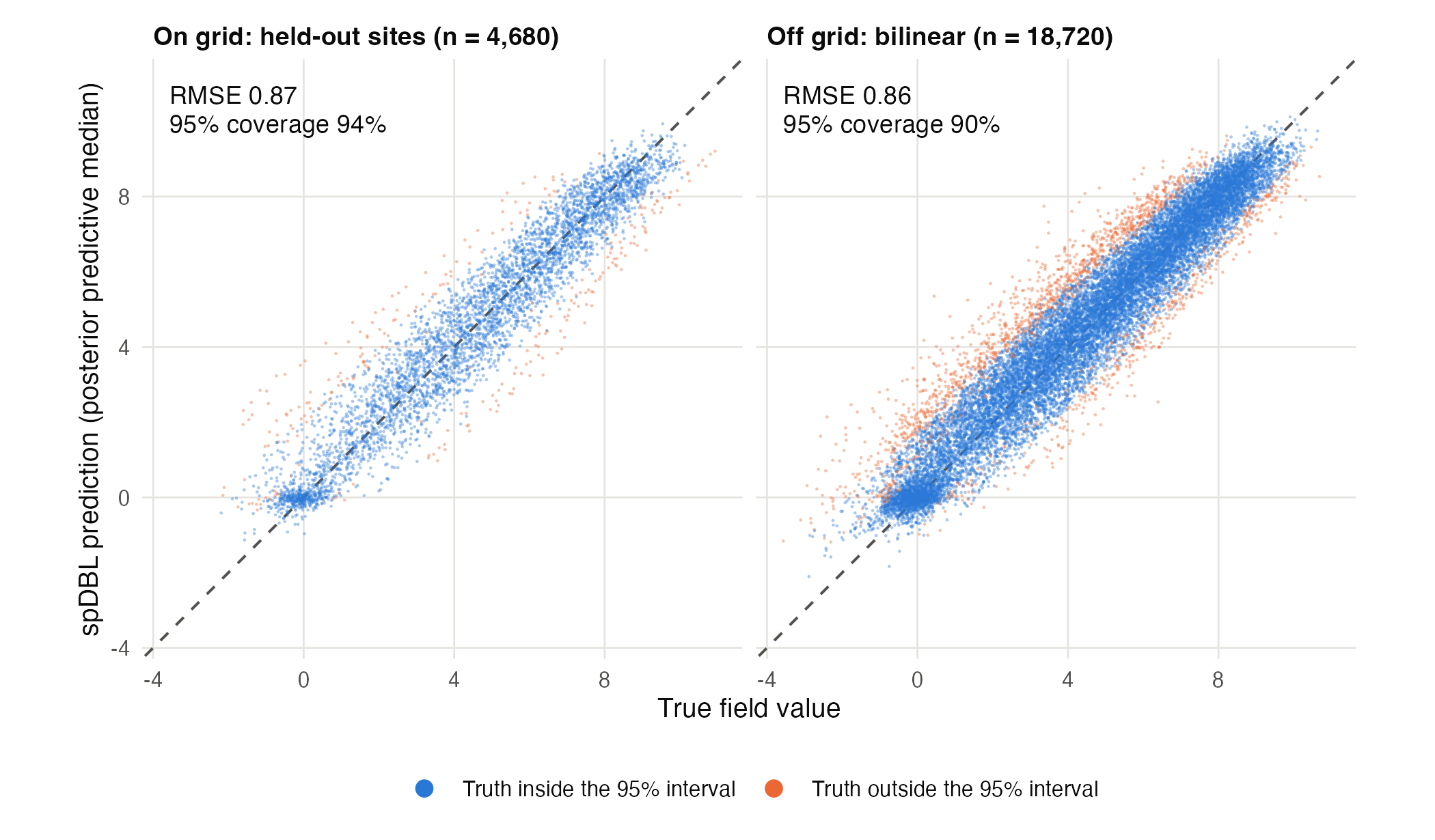}
\caption{Predicted against true values of the field
$y_t(\bm\eta, \bm s_0) + u_t(\bm s_0)$ at locations without field data, in the
30 replicates of E3 at all 26 times (posterior predictive medians; orange: true
value outside the 95\% predictive interval). Left: held-out grid sites; right:
points between cell centres, with the bilinear field map.}
\label{fig:prediction}
\end{figure}

Figure~\ref{fig:prediction} evaluates these predictions as implemented in the
development version of \pkg{spDBL} on GitHub (\code{calibration\_predict()}
with \code{sites} or \code{coords}), in
the 30 replicates of E3 with 100 training runs (10 emulator draws, 2 chains of
1{,}000 iterations). On the grid, 6 of the 36 sites of each replicate were left
out of the calibration and predicted at all 26 times; the emulated field is
drawn as in~\eqref{eq:y-ongrid}, with the field covariance of the calibration,
which adds the uncertainty of the predicted design row $\bm F(\bm\eta)$
(Section~2). Off the grid, the truth was
simulated on an $18\times18$ grid of the same domain, field data were observed
at the 36 coarse cell centres, and 24 points per replicate between them were
predicted. Because the emulator's coefficients are shared by the sites of a
block, \pkg{spDBL} does not impute $\bm\theta_{0:t}(\bm s_0)$ as in Case~2: it
takes the emulated field at $\bm s_0$ from the bilinear interpolation of the
four surrounding cell centres and predicts the discrepancy at $\bm s_0$ itself
as in~\eqref{eq:u-cond}, from the covariance function behind $\bm U$. The
predictive medians had RMSE 0.87 on the grid and 0.86 off the grid, and the
95\% intervals covered 94\% and 90\% of the true values
$y_t(\bm\eta, \bm s_0) + u_t(\bm s_0)$, against 26\% and 20\% for the
calibrated emulator without the discrepancy. Off the grid, coverage was 94\% at
interior points and 76\% within half a coarse cell of the domain's edge, where
the fine solution's absorbing boundary lowers the field below what the coarse
simulator produces. Scripts and results are in the repository's
\code{benchmarks} folder (\code{results/e3/sites} and \code{results/e3/offgrid}).

\section{Analytical WAIC for Matrix-Normal Inverse-Wishart Models}\label{app:waic}

The widely applicable information criterion \citep[WAIC;][]{Watanabe2010}
estimates predictive accuracy from the log pointwise predictive density (lppd)
and an effective number of parameters, here the posterior variance of the
pointwise log likelihood \citep{Gelman2014}. For the conjugate matrix-normal
inverse-Wishart (MNIW) regression both terms have closed forms, which
\code{compute\_analytical\_waic()} evaluates without posterior simulation;
\code{compute\_sample\_waic()} gives the Monte Carlo estimate from posterior
draws. The derivation follows S.~Banerjee's development notes, from which both
functions were adapted.

\subsection{Model and posterior.}
Let $\bm Y$ be an $n\times q$ response matrix with rows $\bm y_i$ and $\bm X$ an
$n\times p$ design matrix with rows $\bm x_i$, and let the rows be independent
observations,
\begin{equation}\label{eq:waic-model}
\bm y_i \mid \bm B, \bm\Sigma \stackrel{\text{ind.}}{\sim} \mathcal{N}_q(\bm x_i\bm B, \bm\Sigma),\quad
\bm B \mid \bm\Sigma \sim \mathcal{MN}_{p\times q}(\bm m_0, \bm M_0, \bm\Sigma),\quad
\bm\Sigma \sim \mathcal{IW}(\nu_0, \bm\Psi_0),
\end{equation}
where the $\mathcal{IW}(\nu, \bm\Psi)$ density is proportional to
$|\bm\Sigma|^{-(\nu+q+1)/2}\exp\{-\operatorname{tr}(\bm\Psi\bm\Sigma^{-1})/2\}$.
With $\mathcal{D} = (\bm Y, \bm X)$, the posterior is
$\bm B \mid \bm\Sigma, \mathcal{D} \sim \mathcal{MN}_{p\times q}(\bm m_n, \bm M_n, \bm\Sigma)$
and $\bm\Sigma \mid \mathcal{D} \sim \mathcal{IW}(\nu_n, \bm\Psi_n)$, where
\begin{align*}
\bm M_n^{-1} &= \bm M_0^{-1} + \bm X^\top\bm X, &
\bm m_n &= \bm M_n(\bm M_0^{-1}\bm m_0 + \bm X^\top\bm Y),\\
\nu_n &= \nu_0 + n, &
\bm\Psi_n &= \bm\Psi_0 + \bm Y^\top\bm Y + \bm m_0^\top\bm M_0^{-1}\bm m_0
  - \bm m_n^\top\bm M_n^{-1}\bm m_n;
\end{align*}
\code{MNIW\_R()} returns them (\code{Vbnew}, \code{Cnew}, \code{vnew},
\code{Snew}) when its argument $\bm H$, the $n\times n$ row covariance of the
errors, is $\bm I_n$, as for the independent rows here.

\subsection{Posterior predictive density.}
For $\tilde m$ new rows $\tilde{\bm Y}$ with design $\tilde{\bm X}$ and row
covariance $\bm\Omega$,
$p(\tilde{\bm Y} \mid \bm B, \bm\Sigma) = \mathcal{MN}_{\tilde m\times q}(\tilde{\bm Y} \mid \tilde{\bm X}\bm B, \bm\Omega, \bm\Sigma)$,
and Bayes' identity gives, for any $(\bm B, \bm\Sigma)$,
\[
p(\tilde{\bm Y} \mid \mathcal{D})
= \frac{p(\tilde{\bm Y} \mid \bm B, \bm\Sigma)\, p(\bm B, \bm\Sigma \mid \mathcal{D})}
       {p(\bm B, \bm\Sigma \mid \mathcal{D}, \tilde{\bm Y})}.
\]
Updating the posterior with $\tilde{\bm Y}$ keeps it MNIW, with
$(\bm M_n^\ast)^{-1} = \bm M_n^{-1} + \tilde{\bm X}^\top\bm\Omega^{-1}\tilde{\bm X}$,
$\nu_n^\ast = \nu_n + \tilde m$ and, by completing the square in $\bm B$ and the
Woodbury identity,
\[
\bm\Psi_n^\ast = \bm\Psi_n + \tilde{\bm E}^\top(\bm\Omega + \tilde{\bm X}\bm M_n\tilde{\bm X}^\top)^{-1}\tilde{\bm E},
\qquad \tilde{\bm E} = \tilde{\bm Y} - \tilde{\bm X}\bm m_n.
\]
Evaluating the identity at the updated mean of $\bm B$, the exponential kernels
and the powers of $|\bm\Sigma|$ cancel, and Sylvester's determinant identity,
$|\bm I_p + \bm M_n\tilde{\bm X}^\top\bm\Omega^{-1}\tilde{\bm X}| = |\bm\Omega|^{-1}|\bm\Omega + \tilde{\bm X}\bm M_n\tilde{\bm X}^\top|$,
collects the determinants into the matrix-$t$ density
\begin{equation}\label{eq:waic-mt}
\begin{split}
p(\tilde{\bm Y} \mid \mathcal{D})
&= \frac{\Gamma_q\{(\nu_n+\tilde m)/2\}}{\Gamma_q(\nu_n/2)\,\pi^{\tilde m q/2}}\,
  |\bm\Omega + \tilde{\bm X}\bm M_n\tilde{\bm X}^\top|^{-q/2}\,|\bm\Psi_n|^{-\tilde m/2}\\
&\qquad\times \left|\bm I_q + \bm\Psi_n^{-1}\tilde{\bm E}^\top
  (\bm\Omega + \tilde{\bm X}\bm M_n\tilde{\bm X}^\top)^{-1}\tilde{\bm E}\right|^{-(\nu_n+\tilde m)/2},
\end{split}
\end{equation}
where $\Gamma_q$ is the multivariate gamma function.

\subsection{Log pointwise predictive density.}
For one row ($\tilde m = 1$, $\tilde{\bm X} = \bm x_i$, $\bm\Omega = 1$), write
$h_i = \bm x_i\bm M_n\bm x_i^\top$, $c_i = 1 + h_i$,
$\bm e_i = \bm y_i - \bm x_i\bm m_n$ and $d_i = \bm e_i\bm\Psi_n^{-1}\bm e_i^\top$.
Since $\Gamma_q\{(\nu_n+1)/2\}/\Gamma_q(\nu_n/2) = \Gamma\{(\nu_n+1)/2\}/\Gamma\{(\nu_n-q+1)/2\}$,
\eqref{eq:waic-mt} is then a $q$-variate Student $t$ density with $\nu_n - q + 1$
degrees of freedom, location $\bm x_i\bm m_n$ and scale $c_i\bm\Psi_n/(\nu_n-q+1)$,
and
\begin{equation}\label{eq:waic-lppd}
\begin{split}
\mathrm{lppd} = \sum_{i=1}^n \bigg[ &\log\Gamma\Big(\frac{\nu_n+1}{2}\Big)
  - \log\Gamma\Big(\frac{\nu_n-q+1}{2}\Big) - \frac q2\log\pi\\
  &- \frac q2\log c_i - \frac12\log|\bm\Psi_n|
  - \frac{\nu_n+1}{2}\log\Big(1 + \frac{d_i}{c_i}\Big)\bigg].
\end{split}
\end{equation}

\subsection{WAIC penalty.}
The penalty is
\[
p_{\mathrm{WAIC}} = \sum_{i=1}^n \operatorname{Var}_{\bm B, \bm\Sigma \mid \mathcal{D}}\{\log p(\bm y_i \mid \bm B, \bm\Sigma)\},
\]
with $\log p(\bm y_i \mid \bm B, \bm\Sigma) = -\frac q2\log(2\pi) - \frac12 A - \frac12 B_i$,
$A = \log|\bm\Sigma|$ and
$B_i = (\bm y_i - \bm x_i\bm B)\bm\Sigma^{-1}(\bm y_i - \bm x_i\bm B)^\top$, so that
each term is $\{\operatorname{Var}(A) + \operatorname{Var}(B_i) + 2\operatorname{Cov}(A, B_i)\}/4$.
Given $\bm\Sigma$, $\bm y_i - \bm x_i\bm B \sim \mathcal{N}_q(\bm e_i, h_i\bm\Sigma)$,
so $B_i/h_i$ is non-central $\chi^2_q$ with non-centrality $w_i/h_i$, where
$w_i = \bm e_i\bm\Sigma^{-1}\bm e_i^\top$, and
$\operatorname{E}(B_i \mid \bm\Sigma) = q h_i + w_i$,
$\operatorname{Var}(B_i \mid \bm\Sigma) = 2q h_i^2 + 4h_i w_i$.
Under $\bm\Sigma^{-1} \mid \mathcal{D} \sim \mathcal{W}_q(\nu_n, \bm\Psi_n^{-1})$,
$w_i/d_i \sim \chi^2_{\nu_n}$,
$\operatorname{Var}(A) = \sum_{j=1}^q \psi_1\{(\nu_n-j+1)/2\}$ with $\psi_1$ the
trigamma function, and $\operatorname{Cov}(\log|\bm\Sigma^{-1}|, w_i) = 2d_i$.
Because $A$ depends on $\bm\Sigma$ alone,
$\operatorname{Cov}(A, B_i) = \operatorname{Cov}\{A, \operatorname{E}(B_i \mid \bm\Sigma)\} = -2d_i$,
and the law of total variance gives
$\operatorname{Var}(B_i) = 2q h_i^2 + 4h_i\nu_n d_i + 2\nu_n d_i^2$. Hence
\begin{equation}\label{eq:waic-penalty}
p_{\mathrm{WAIC}} = \frac14\sum_{i=1}^n\bigg[\sum_{j=1}^q \psi_1\Big(\frac{\nu_n-j+1}{2}\Big)
  + 2q h_i^2 + 4h_i\nu_n d_i + 2\nu_n d_i^2 - 4d_i\bigg],
\end{equation}
and $\mathrm{WAIC} = -2(\mathrm{lppd} - p_{\mathrm{WAIC}})$, lower WAIC
indicating higher estimated predictive accuracy.

\subsection{Implementation.}
\code{compute\_analytical\_waic(Y, X, Mn, mn, Psi\_n, nu\_n)} evaluates
\eqref{eq:waic-lppd} and \eqref{eq:waic-penalty} from Cholesky factors of
$\bm M_n$ and $\bm\Psi_n$: $h_i$ is the squared norm of $\bm x_i$ times a factor
of $\bm M_n$, which keeps it non-negative, $d_i$ the squared norm of the
whitened residual, and $\log(1 + \cdot)$ is computed with \code{log1p()}. It
returns the pointwise and total lppd, $p_{\mathrm{WAIC}}$, expected log
predictive density and WAIC; $q = 1$ is the normal--inverse-gamma case.
\code{compute\_sample\_waic()} computes the same quantities from paired posterior
draws of $(\bm B, \bm\Sigma)$, with a log-sum-exp lppd and sample variances
(denominator $S - 1$, as in \code{loo::waic()}), and can return the
$S \times n$ log-likelihood matrix for the \pkg{loo} package. The tests check the
analytical scores against Monte Carlo estimates from independent MNIW draws and
the sampled scores against \pkg{loo}'s aggregation, and
\code{vignette("model-comparison")} compares covariate subsets. Each row of
$\bm Y$ is one observation, so the formulas score models with independent rows;
the emulator in~\eqref{eq:emulator}, whose rows are correlated through
$\bm V$, needs its own factorisation of the likelihood.

\end{document}

%% file: tables/appendix-tables.tex
\begin{table}[ht]
\centering\footnotesize\setlength{\tabcolsep}{4pt}
\begin{tabular}{@{}rr*{3}{rrr}@{}}
\toprule
 & & \multicolumn{3}{c}{$w = 1$} & \multicolumn{3}{c}{$w = 2$} & \multicolumn{3}{c}{$w = 5$} \\
\cmidrule(lr){3-5}\cmidrule(lr){6-8}\cmidrule(lr){9-11}
Grid & $S$ & fit & predict & peak & fit & predict & peak & fit & predict & peak \\
\midrule
$10^2$ & 100 & 0.18 & 0.05 & 0.2/0.2 & 0.20 & 0.04 & 0.2/0.2 & 0.50 & 0.04 & 0.2/0.2 \\
$20^2$ & 400 & 0.92 & 0.16 & 0.3/0.2 & 1.61 & 0.14 & 0.4/0.3 & 4.01 & 0.14 & 0.7/0.4 \\
$30^2$ & 900 & 2.52 & 0.28 & 0.7/0.4 & 4.71 & 0.28 & 1.0/0.5 & 13.8 & 0.33 & 1.7/1.0 \\
$40^2$ & 1,600 & 5.30 & 0.46 & 1.1/0.7 & 11.2 & 0.52 & 1.9/1.1 & 35.3 & 0.63 & 3.8/2.2 \\
$50^2$ & 2,500 & 10.1 & 0.72 & 1.8/1.2 & 22.0 & 0.77 & 3.6/1.8 & 76.4 & 1.06 & 8.2/4.0 \\
$70^2$ & 4,900 & 27.6 & 1.42 & 5.4/3.3 & 64.1 & 1.38 & 9.6/5.3 & 252 & 2.46 & 21.8/10.6 \\
$100^2$ & 10,000 & 86.6 & 3.18 & 13.3/9.1 & 214 & 3.82 & 23.3/15.5 & 929 & 5.88 & 61.6/33.5 \\
\bottomrule
\end{tabular}
\caption{E1: inverse-Wishart emulator with 10 posterior draws, 40 training runs and 26 time
steps; fit and prediction (10 held-out runs) time in seconds and peak memory of the
whole task in GiB, one task per node, with stored coefficient draws and, after the slash,
with draws regenerated on demand (\code{emulator\_learn(keep\_draws = FALSE)}; same
accuracy). Times are with stored draws.}
\label{tab:e1}
\end{table}

\begin{table}[ht]
\centering\scriptsize\setlength{\tabcolsep}{4pt}
\begin{tabular}{@{}llrrrrrrr@{}}
\toprule
Grid & Method & RMSE & CRPS & Cov. & Width & Fit (s) & Predict (s) & Peak (GiB) \\
\midrule
$10^2$ & \pkg{spDBL}, tuned$^{d}$ & 0.453 & 0.125 & 0.961 & 0.92 & 9.25 & 1.39 & 0.36 \\
 & \code{ppgasp}, tuned$^{d}$ & 0.432 & 0.106 & 0.924 & 0.59 & 3.19 & -- & 0.17 \\
 & PCA-GP, tuned$^{d}$ & 0.415 & 0.124 & 0.839 & 0.46 & 7.75 & 0.13 & 0.14 \\
 & \pkg{spDBL}, design GP & 0.454 & 0.127 & 0.958 & 0.92 & 8.09 & 0.88 & 0.23 \\
 & design GP alone$^{a}$ & 0.467 & 0.127 & 0.949 & 0.88 & 8.45 & 0.47 & 0.23 \\
 & \pkg{spDBL}, known AR rows$^{b}$ & 0.068 & 0.024 & 0.986 & 0.28 & 0.31 & 0.25 & 0.21 \\
 & \code{ppgasp}, default & 0.474 & 0.113 & 0.857 & 0.40 & 0.58 & 0.01 & 0.14 \\
 & \code{ppgasp}, nugget, Nelder--Mead & 0.466 & 0.113 & 0.867 & 0.42 & 2.86 & 0.01 & 0.14 \\
 & \code{ppgasp} per time step & 0.457 & 0.106 & 0.855 & 0.40 & 3.91 & -- & 0.17 \\
 & PCA-GP (\pkg{mlegp}) & 0.430 & 0.133 & 0.819 & 0.56 & 0.78 & 0.08 & 0.14 \\
 & \pkg{spDBL}, design GP$^{c}$ & 0.454 & 0.127 & 0.958 & 0.92 & 8.06 & 1.09 & 0.21 \\
\addlinespace
$20^2$ & \pkg{spDBL}, tuned$^{d}$ & 0.612 & 0.186 & 0.955 & 1.31 & 18.7 & 5.25 & 1.74 \\
 & \code{ppgasp}, tuned$^{d}$ & 0.592 & 0.157 & 0.920 & 0.85 & 7.73 & -- & 0.21 \\
 & PCA-GP, tuned$^{d}$ & 0.597 & 0.193 & 0.861 & 0.85 & 9.06 & 0.16 & 0.18 \\
 & \pkg{spDBL}, design GP & 0.613 & 0.187 & 0.951 & 1.31 & 10.5 & 2.67 & 0.58 \\
 & design GP alone$^{a}$ & 0.630 & 0.189 & 0.949 & 1.31 & 10.4 & 0.92 & 0.53 \\
 & \pkg{spDBL}, known AR rows$^{b}$ & 0.099 & 0.028 & 0.980 & 0.31 & 2.08 & 0.99 & 0.52 \\
 & \code{ppgasp}, default & 0.682 & 0.178 & 0.859 & 0.64 & 2.47 & 0.02 & 0.18 \\
 & \code{ppgasp}, nugget, Nelder--Mead & 0.663 & 0.177 & 0.868 & 0.67 & 10.5 & 0.03 & 0.17 \\
 & \code{ppgasp} per time step & 0.661 & 0.167 & 0.861 & 0.63 & 5.62 & -- & 0.20 \\
 & PCA-GP (\pkg{mlegp}) & 0.593 & 0.205 & 0.854 & 0.89 & 0.92 & 0.10 & 0.18 \\
 & \pkg{spDBL}, design GP$^{c}$ & 0.613 & 0.187 & 0.951 & 1.31 & 9.53 & 4.01 & 0.42 \\
\addlinespace
$50^2$ & \pkg{spDBL}, tuned$^{d}$ & 0.780 & 0.259 & 0.948 & 1.79 & 191 & 42.2 & 20.42 \\
 & \code{ppgasp}, tuned$^{d}$ & 0.735 & 0.212 & 0.918 & 1.16 & 39.6 & -- & 0.63 \\
 & PCA-GP, tuned$^{d}$ & 0.753 & 0.261 & 0.873 & 1.13 & 10.2 & 0.24 & 0.59 \\
 & \pkg{spDBL}, design GP & 0.782 & 0.260 & 0.945 & 1.79 & 39.9 & 16.1 & 5.34 \\
 & design GP alone$^{a}$ & 0.797 & 0.262 & 0.946 & 1.81 & 39.5 & 3.22 & 4.65 \\
 & \pkg{spDBL}, known AR rows$^{b}$ & 0.225 & 0.041 & 0.969 & 0.41 & 29.5 & 6.70 & 5.32 \\
 & \code{ppgasp}, default & 0.918 & 0.262 & 0.867 & 0.99 & 9.81 & 0.20 & 0.62 \\
 & \code{ppgasp}, nugget, Nelder--Mead & 0.859 & 0.255 & 0.884 & 1.12 & 64.9 & 0.20 & 0.61 \\
 & \code{ppgasp} per time step & 0.883 & 0.244 & 0.876 & 0.99 & 19.8 & -- & 0.68 \\
 & PCA-GP (\pkg{mlegp}) & 0.769 & 0.281 & 0.863 & 1.20 & 1.23 & 0.18 & 0.60 \\
 & \pkg{spDBL}, design GP$^{c}$ & 0.782 & 0.260 & 0.945 & 1.79 & 29.5 & 36.9 & 2.41 \\
\bottomrule
\end{tabular}
\caption{E2 on the fixed $10\times 10$ domain: held-out prediction on 50 runs at times
\mbox{$t \ge 3$}, means over 10 independent training/test designs; Cov.\ and Width are the coverage
and mean width of central 95\% intervals (\pkg{spDBL}'s from \code{predict(..., interval
= "prediction")}). Fit and prediction time and peak memory are from design 1, rerun with
spDBL 1.2.0's batched prediction, one task per node; \code{ppgasp} per time step fits one \code{ppgasp} per time, and its fit time includes
prediction. $^{a}$Times $3 \le t \le 25$: it has no prediction for the last time, $t = 26$. $^{b}$Uses the
simulator's outputs at the test inputs to build the autoregressive rows (one-step prediction), so
it is not simulator-free. $^{c}$With coefficient draws regenerated on demand
(\code{emulator\_learn(keep\_draws = FALSE)}); its predictions are identical.
$^{d}$Each method's setting selected on designs 11--12 (Table~\ref{tab:e2-tuning}): \pkg{spDBL} with episode
window 5, \code{gp\_tune = 1} and 50 draws; \code{ppgasp} fitted per time step with a linear trend and
the Mat\'ern 3/2 kernel; the PCA-GP with a linear mean, 99.99\% of variance, a nugget and 20
restarts; times and memory from design 1, one task per node; the tuned \code{ppgasp}'s fit time includes prediction.}
\label{tab:e2}
\end{table}

\begin{table}[ht]
\centering\footnotesize\setlength{\tabcolsep}{3pt}
\begin{tabular}{@{}llrrr@{}}
\toprule
Grid & Method & $\Delta$ RMSE & $\Delta$ CRPS & $\Delta$ coverage (points) \\
\midrule
$10^2$ & \code{ppgasp}, tuned & $-$0.021 [$-$0.047, 0.005] & $-$0.019 [$-$0.028, $-$0.010] & $-$3.7 [$-$4.8, $-$2.6] \\
 & PCA-GP, tuned & $-$0.038 [$-$0.090, 0.014] & $-$0.001 [$-$0.030, 0.027] & $-$12.2 [$-$14.5, $-$9.8] \\
\addlinespace
$20^2$ & \code{ppgasp}, tuned & $-$0.020 [$-$0.068, 0.028] & $-$0.029 [$-$0.044, $-$0.014] & $-$3.5 [$-$4.5, $-$2.5] \\
 & PCA-GP, tuned & $-$0.014 [$-$0.088, 0.059] & 0.007 [$-$0.022, 0.036] & $-$9.4 [$-$12.7, $-$6.0] \\
\addlinespace
$50^2$ & \code{ppgasp}, tuned & $-$0.045 [$-$0.091, $-$0.0003] & $-$0.047 [$-$0.066, $-$0.028] & $-$3.0 [$-$4.0, $-$2.0] \\
 & PCA-GP, tuned & $-$0.028 [$-$0.099, 0.043] & 0.002 [$-$0.024, 0.029] & $-$7.5 [$-$9.5, $-$5.6] \\
\bottomrule
\end{tabular}
\caption{E2 with every method's tuned setting (Table~\ref{tab:e2}, note d): each competitor minus
tuned \pkg{spDBL}, paired by design, mean and 95\% $t$ interval over the 10 designs, at times $t \ge 3$.
Positive RMSE and CRPS differences favour \pkg{spDBL}; negative coverage differences mean the
competitor's 95\% intervals cover less.}
\label{tab:e2-paired}
\end{table}

\begin{table}[ht]
\centering\footnotesize\setlength{\tabcolsep}{4pt}
\begin{tabular}{@{}llrrrr@{}}
\toprule
Method & Setting & Round & CRPS & RMSE & Cov. \\
\midrule
\pkg{spDBL} & \textbf{window 5, \code{gp\_tune} 1, 50 draws} & 2 & 0.1182 & 0.432 & 0.959 \\
 & window 5, \code{gp\_tune} 1 & 2 & 0.1183 & 0.432 & 0.959 \\
 & window 5, 50 draws & 2 & 0.1184 & 0.432 & 0.958 \\
 & episode window 5 & 1 & 0.1184 & 0.432 & 0.958 \\
 & \code{gp\_tune} 1, 50 draws & 2 & 0.1199 & 0.433 & 0.958 \\
 & \code{gp\_tune} 1 & 1 & 0.1200 & 0.433 & 0.958 \\
 & 50 draws & 1 & 0.1200 & 0.433 & 0.955 \\
 & E2 setting: window 2, \code{gp\_tune} 0.5, 20 draws & 1 & 0.1200 & 0.433 & 0.955 \\
 & \code{gp\_tau2} $10^{-2}$ & 1 & 0.1201 & 0.433 & 0.955 \\
 & AR(1) design & 1 & 0.1204 & 0.434 & 0.955 \\
 & \code{gp\_tune} 0.25 & 1 & 0.1209 & 0.434 & 0.955 \\
 & episode window 1 & 1 & 0.1216 & 0.435 & 0.953 \\
 & \code{gp\_sigma2} 4 & 1 & 0.1226 & 0.434 & 0.960 \\
 & squared-exponential design GP & 1 & 0.1250 & 0.446 & 0.938 \\
 & one design GP for all times & 1 & 0.1411 & 0.461 & 0.961 \\
\addlinespace
\code{ppgasp} & \textbf{per time step, trend, Mat\'ern 3/2} & 2 & 0.1058 & 0.442 & 0.914 \\
 & per time step, trend, power exp. & 2 & 0.1060 & 0.442 & 0.928 \\
 & per time step, Mat\'ern 3/2 & 2 & 0.1072 & 0.442 & 0.915 \\
 & per time step, power exp. & 2 & 0.1108 & 0.445 & 0.935 \\
 & per time step, trend & 1 & 0.1139 & 0.493 & 0.841 \\
 & per time step & 1 & 0.1147 & 0.493 & 0.840 \\
 & trend, Mat\'ern 3/2 & 2 & 0.1171 & 0.461 & 0.913 \\
 & Mat\'ern 3/2 & 1 & 0.1240 & 0.473 & 0.911 \\
 & power exponential & 1 & 0.1302 & 0.481 & 0.932 \\
 & linear trend & 1 & 0.1336 & 0.533 & 0.835 \\
 & 10 optimiser starts & 1 & 0.1369 & 0.542 & 0.833 \\
 & \code{mmle} & 1 & 0.1369 & 0.542 & 0.833 \\
 & \code{ref\_gamma} prior & 1 & 0.1369 & 0.542 & 0.833 \\
 & nugget, Nelder--Mead (E2's tuned) & 1 & 0.1369 & 0.542 & 0.833 \\
\addlinespace
PCA-GP & \textbf{linear mean, 99.99\%, nugget, 20 restarts} & 2 & 0.1180 & 0.432 & 0.837 \\
 & linear mean, 99.99\%, nugget & 2 & 0.1181 & 0.431 & 0.841 \\
 & linear mean, nugget & 2 & 0.1205 & 0.432 & 0.808 \\
 & linear mean, 99.99\% & 2 & 0.1230 & 0.436 & 0.825 \\
 & 99.99\%, nugget & 2 & 0.1236 & 0.437 & 0.844 \\
 & linear mean & 1 & 0.1250 & 0.438 & 0.792 \\
 & 99.99\% of variance & 1 & 0.1255 & 0.439 & 0.844 \\
 & nugget & 1 & 0.1258 & 0.439 & 0.813 \\
 & 20 restarts & 1 & 0.1275 & 0.440 & 0.809 \\
 & E2 setting: 99.9\% of variance & 1 & 0.1276 & 0.440 & 0.810 \\
 & 99\% of variance & 1 & 0.1428 & 0.445 & 0.655 \\
\bottomrule
\end{tabular}
\caption{E2 tuning study: every setting tried for each method on designs 11 and 12, which E2 does not
evaluate, all three grids, means over the six grid--design pairs, sorted by mean CRPS. Round 1 changes
one setting at a time from the method's E2 setting (the row labelled E2); round 2 combines the round-1
changes that lowered CRPS. The setting with the lowest mean CRPS (bold; ties broken by RMSE), a rule
fixed before any result, is the tuned setting of Tables~\ref{tab:e2} and~\ref{tab:e2-paired}. Settings
not listed are the package defaults. Three \code{ppgasp} settings (10 starts, \code{mmle},
\code{ref\_gamma}) change its mean CRPS by under $10^{-5}$.}
\label{tab:e2-tuning}
\end{table}

\begin{table}[ht]
\centering\scriptsize\setlength{\tabcolsep}{3pt}
\begin{tabular}{@{}llrrrrr@{}}
\toprule
& & & & & \multicolumn{2}{c}{Method minus \pkg{spDBL}} \\ \cmidrule(l){6-7}
Method & Parameter & Relative RMSE & Coverage & Rel.\ width & Relative RMSE & Coverage \\
\midrule
\pkg{spDBL} & $\eta_1$ & 0.038 [0.029, 0.045] & 0.73 [0.57, 0.90] & 0.09 & -- & -- \\
 & $\eta_2$ & 0.140 [0.111, 0.171] & 0.87 [0.73, 0.97] & 0.37 & -- & -- \\
 & $\alpha_2$ & 0.037 [0.029, 0.045] & 0.80 [0.67, 0.93] & 0.11 & -- & -- \\
 & all & 0.087 [0.071, 0.103] & 0.80 [0.70, 0.89] & 0.19 & -- & -- \\
\addlinespace
S-GaSP & $\eta_1$ & 0.039 [0.031, 0.047] & 0.73 [0.57, 0.87] & 0.09 & 0.002 [$-$0.008, 0.012] & 0.00 [$-$0.20, 0.20] \\
 & $\eta_2$ & 0.169 [0.133, 0.202] & 0.50 [0.33, 0.67] & 0.21 & 0.028 [$-$0.012, 0.068] & $-$0.37 [$-$0.57, $-$0.17] \\
 & $\alpha_2$ & 0.040 [0.029, 0.051] & 0.77 [0.60, 0.90] & 0.11 & 0.003 [$-$0.008, 0.013] & $-$0.03 [$-$0.13, 0.07] \\
 & all & 0.103 [0.084, 0.121] & 0.67 [0.56, 0.78] & 0.14 & 0.016 [$-$0.006, 0.037] & $-$0.13 [$-$0.26, $-$0.01] \\
\addlinespace
GaSP & $\eta_1$ & 0.037 [0.029, 0.045] & 0.80 [0.63, 0.93] & 0.09 & 0.000 [$-$0.009, 0.009] & 0.07 [$-$0.13, 0.27] \\
 & $\eta_2$ & 0.151 [0.114, 0.184] & 0.67 [0.50, 0.83] & 0.30 & 0.010 [$-$0.030, 0.048] & $-$0.20 [$-$0.40, 0.03] \\
 & $\alpha_2$ & 0.038 [0.027, 0.049] & 0.80 [0.63, 0.93] & 0.11 & 0.001 [$-$0.009, 0.011] & 0.00 [$-$0.10, 0.10] \\
 & all & 0.092 [0.073, 0.111] & 0.76 [0.64, 0.86] & 0.17 & 0.006 [$-$0.016, 0.026] & $-$0.04 [$-$0.16, 0.07] \\
\bottomrule
\end{tabular}
\caption{E3 with 100 Latin-hypercube training runs (one design for every replicate and method): 30 replicate calibrations of the three free inputs $\eta = (\eta_1, \eta_2, \alpha_2)$ by \pkg{spDBL} and by \pkg{RobustCalibration} with its S-GaSP and GaSP discrepancies, every chain started at its package's optimum from 20 starts (\pkg{spDBL}: conditional posterior mode from 20 L-BFGS starts, 20 emulator draws with 4 NUTS chains of 2{,}000 iterations each; \pkg{RobustCalibration}: maximum-likelihood estimate from 20 starts, an emulator nugget, random-walk proposal scales set from a 4{,}000-iteration pilot chain, and 4 chains of 20{,}000 iterations); errors and widths relative to the prior range, with replicate-bootstrap 95\% intervals, paired over replicates for the differences. Replicates with rank-normalised $\hat R$ above 1.01 and above 1.1 for some $\eta$ component (for \pkg{spDBL} the worst of its 20 conditional fits, for \pkg{RobustCalibration} its four chains): \pkg{spDBL} 10 and 1 of 30, S-GaSP 1 and 1 of 30, GaSP 3 and 1 of 30. Mean CPU hours per calibration, one Slurm job per calibration (\pkg{spDBL} on one core; \pkg{RobustCalibration} running its four chains on four cores, CPU time summed over them and including the mode search and pilot): \pkg{spDBL} 9.4, S-GaSP 16.8, GaSP 10.3.}
\label{tab:e3}
\end{table}

\begin{table}[ht]
\centering\scriptsize\setlength{\tabcolsep}{3pt}
\begin{tabular}{@{}llrrrrr@{}}
\toprule
& & & & & \multicolumn{2}{c}{Method minus \pkg{spDBL}} \\ \cmidrule(l){6-7}
Method & Parameter & Relative RMSE & Coverage & Rel.\ width & Relative RMSE & Coverage \\
\midrule
\pkg{spDBL} & $\eta_1$ & 0.053 [0.040, 0.064] & 0.43 [0.27, 0.60] & 0.06 & -- & -- \\
 & $\eta_2$ & 0.156 [0.105, 0.201] & 0.57 [0.40, 0.73] & 0.21 & -- & -- \\
 & $\alpha_2$ & 0.050 [0.036, 0.063] & 0.63 [0.47, 0.80] & 0.09 & -- & -- \\
 & all & 0.100 [0.073, 0.123] & 0.54 [0.42, 0.67] & 0.12 & -- & -- \\
\addlinespace
S-GaSP & $\eta_1$ & 0.043 [0.036, 0.051] & 0.70 [0.53, 0.83] & 0.09 & $-$0.009 [$-$0.022, 0.005] & 0.27 [0.03, 0.50] \\
 & $\eta_2$ & 0.176 [0.139, 0.208] & 0.43 [0.27, 0.60] & 0.17 & 0.019 [$-$0.038, 0.079] & $-$0.13 [$-$0.37, 0.10] \\
 & $\alpha_2$ & 0.044 [0.033, 0.054] & 0.67 [0.50, 0.83] & 0.10 & $-$0.006 [$-$0.023, 0.009] & 0.03 [$-$0.17, 0.27] \\
 & all & 0.107 [0.088, 0.125] & 0.60 [0.48, 0.72] & 0.12 & 0.008 [$-$0.023, 0.039] & 0.06 [$-$0.10, 0.21] \\
\addlinespace
GaSP & $\eta_1$ & 0.037 [0.030, 0.043] & 0.77 [0.60, 0.90] & 0.09 & $-$0.016 [$-$0.028, $-$0.002] & 0.33 [0.10, 0.57] \\
 & $\eta_2$ & 0.198 [0.146, 0.248] & 0.50 [0.33, 0.67] & 0.26 & 0.041 [$-$0.011, 0.090] & $-$0.07 [$-$0.30, 0.17] \\
 & $\alpha_2$ & 0.039 [0.029, 0.049] & 0.73 [0.57, 0.87] & 0.10 & $-$0.011 [$-$0.027, 0.005] & 0.10 [$-$0.10, 0.30] \\
 & all & 0.118 [0.090, 0.147] & 0.67 [0.57, 0.77] & 0.15 & 0.019 [$-$0.010, 0.045] & 0.12 [$-$0.02, 0.27] \\
\bottomrule
\end{tabular}
\caption{E3 with the 40 training runs bundled with the package (\code{dt\_calibration}), with \pkg{RobustCalibration}'s default proposal scales, no emulator nugget and 10 maximum-likelihood starts: every chain started at its package's optimum (\pkg{spDBL}: conditional posterior mode, \code{init = "mode"}, 20 emulator draws with 4 NUTS chains of 2{,}000 iterations each; \pkg{RobustCalibration}: its 10-start maximum-likelihood estimate, 4 chains of 20{,}000 iterations); errors and widths relative to the prior range, with replicate-bootstrap 95\% intervals, paired over replicates for the differences. Replicates with rank-normalised $\hat R$ above 1.01 and above 1.1 for some $\eta$ component (for \pkg{spDBL} the worst of its 20 conditional fits, for \pkg{RobustCalibration} its four chains): \pkg{spDBL} 13 and 5 of 30, S-GaSP 30 and 4 of 30, GaSP 30 and 14 of 30. Mean CPU hours per calibration, with 18 one-core tasks per 36-core node: \pkg{spDBL} 8.4, S-GaSP 14.5, GaSP 9.0.}
\label{tab:e3-40}
\end{table}

\begin{table}[ht]
\centering\scriptsize\setlength{\tabcolsep}{3pt}
\begin{tabular}{@{}lrrrrrr@{}}
\toprule
& \multicolumn{2}{c}{Coverage} & Relative & \multicolumn{2}{c}{Relative RMSE} & $\hat R > 1.01$ \\
\cmidrule(lr){2-3}\cmidrule(lr){5-6}
Variant & value & minus baseline & width & value & minus baseline & replicates \\
\midrule
Baseline (Table~\ref{tab:e3-40}) & 0.57 & -- & 0.123 & 0.086 & -- & 6 \\
Joint paths & 0.53 & $-$0.03 [$-$0.10, 0.00] & 0.123 & 0.087 & 0.001 [0.000, 0.002] & 4 \\
60 draws & 0.53 & $-$0.03 [$-$0.10, 0.00] & 0.125 & 0.086 & 0.000 [$-$0.001, 0.001] & 6 \\
No design uncertainty & 0.77 & 0.20 [0.00, 0.37] & 0.202 & 0.084 & $-$0.002 [$-$0.027, 0.025] & 3 \\
Data from emulator model & 0.93 & 0.37 [0.17, 0.57] & 0.182 & 0.045 & $-$0.041 [$-$0.081, $-$0.003] & 6 \\
New 40-run design & 0.77 & 0.20 [$-$0.03, 0.43] & 0.186 & 0.061 & $-$0.025 [$-$0.065, 0.015] & 0 \\
80 training runs & 0.83 & 0.27 [0.07, 0.50] & 0.215 & 0.080 & $-$0.005 [$-$0.030, 0.021] & 2 \\
100 training runs & 0.83 & 0.27 [0.07, 0.47] & 0.191 & 0.088 & 0.002 [$-$0.031, 0.038] & 2 \\
160 training runs & 0.93 & 0.37 [0.20, 0.53] & 0.220 & 0.069 & $-$0.017 [$-$0.056, 0.019] & 2 \\
40 + 40 targeted runs & 1.00 & 0.43 [0.23, 0.63] & 0.221 & 0.052 & $-$0.033 [$-$0.071, 0.001] & 0 \\
\bottomrule
\end{tabular}
\caption{E3 ablations: \pkg{spDBL} on replicates 1--10 with one change each from the setting of Table~\ref{tab:e3-40}, against that run on the same replicates and seeds, with paired replicate-bootstrap 95\% intervals of the differences. Joint paths: \code{calibration\_prepare(sampling = "joint")}. 60 draws: 60 emulator draws instead of 20. No design uncertainty: \code{design\_uncertainty = FALSE}. New 40-run design, 80, 100 and 160 training runs: new Latin-hypercube designs over the prior box instead of the 40 bundled runs (100 runs: replicates 1--10 of Table~\ref{tab:e3}). 40 + 40 targeted runs: the bundled runs plus 40 placed by greedy maximin where a short first-stage calibration puts the posterior, its spread tripled. Data from emulator model: the simulator output in each replicate's field data replaced by a draw from the emulator's own predictive model at the true parameters (one random emulator draw; the same discrepancy and noise), so that \pkg{spDBL} is exactly specified.}
\label{tab:ablation-e3}
\end{table}